\documentclass[%
pra,
twocolumn,
reprint,
amsmath,amssymb,
aps,
]{revtex4-1}
\usepackage{graphicx}
\usepackage{dcolumn}
\usepackage{bm}

\usepackage{theorem}
\usepackage{makeidx}
\theorembodyfont{\normalfont}
\newtheorem{theo}{Theorem}

\newtheorem{lemm}{Lemma}

\begin{document}

\preprint{APS/123-QED}
\title{$\mu$-Symmetry breaking: an algebraic approach\\ to finding mean fields of quantum many-body systems}
\author{Sho Higashikawa${}^1$}
\author{Masahito Ueda${}^{1,2}$}
\affiliation{
${}^1$ Department of Physics, University of Tokyo, 7-3-1 Hongo, Bunkyo-ku, Tokyo 113-0033, Japan \\
${}^2$ RIKEN Center for Emergent Matter Science (CEMS), Wako, Saitama 351-0198, Japan
}
\date{\today}
\begin{abstract}
One of the most fundamental problems in quantum many-body systems is the identification of a mean field in spontaneous symmetry breaking 
which is usually made in a heuristic manner.
We propose a systematic method of finding a mean field based on the Lie algebra and the dynamical symmetry 
by introducing a class of symmetry broken phases which we call $\mu$-symmetry breaking. 
We show that for $\mu$-symmetry breaking
the quadratic part of an effective Lagrangian of Nambu-Goldstone modes can be block-diagonalized  
and that homotopy groups of topological excitations can be calculated systematically. 
\end{abstract}
\pacs{Valid PACS appear here}
\maketitle

\newcommand{\Gg}{\mathfrak{g}}
\newcommand{\tbr}[1]{\{#1\}}
\newcommand{\hmustate}{\left| \bm{\mu}_H \right)}
\newcommand{\hmu}{\bm{\mu}_H}
\newcommand{\zmustate}{\left| \bm{\mu}_0 \right)}
\newcommand{\zmu}{\bm{\mu}_0}
\newcommand{\Gh}{\mathfrak{h}}
\newcommand{\mr}[1]{\mathrm{#1}}
\newcommand{\mbZ}{\mathbb{Z}}
\newcommand{\Bal}{\bm{\alpha}}
\newcommand{\vev}{\langle \bm{\phi} \rangle}

\section{Introduction}
Spontaneous symmetry breaking (SSB) has long played a pivotal role in our understanding of Nature \cite{Nambu60}.
Examples include ferromagnetism \cite{Weiss07}, superconductivity \cite{Bardeen57}, Bose-Einstein condensation \cite{Gross61, Pitaevskii61}, 
chiral symmetry breaking \cite{Nambu61a, Nambu61b}, and unification of the fundamental forces \cite{Weinberg67}.
Both static and dynamic properties of a symmetry broken phase can be described by the corresponding mean field, which is usually found in a heuristic manner. 
The identification of the mean field amounts to that of an order parameter 
or that of an operator that supports a long-range order (LRO) in quantum field theory \cite{Yang62}.

In this paper, 
we propose a systematic method of finding mean fields of quantum many-body systems based on the Lie algebra and the dynamical symmetry. 
The dynamical symmetry has achieved a remarkable success in few-body systems for finding the atomic spectrum of hydrogen \cite{Pauli26, Iachello06} 
and collective excitation spectra of nuclei \cite{Iachello84}. 
Here we apply the dynamical symmetry to a particular class of broken symmetry systems in which 
the mean fields are described in terms of the weight vector in the representation of the Lie algebra. 
Since the weight of the Lie algebra is often labeled by the Greek letter $\mu$, 
we refer to such symmetry breaking as $\mu$-symmetry breaking. 
We show that for $\mu$-symmetry breaking 
the quadratic part of an effective Lagrangian of Nambu-Goldstone modes (NG modes) can be block-diagonalized 
and that homotopy groups of topological excitations can be calculated systematically. 
By applying this method to a $U(N)$-symmetric system which has recently been realized in an ultracold atomic gas \cite{Fukuhara07, Taie12}, 
we show that a large class of symmetry broken phases can be described in terms of $\mu$-symmetry breaking. 

This paper is organized as follows. 
In Sec.~\ref{sec:definition}, 
we introduce the concept of $\mu$-symmetry breaking and derive mean fields 
by combining it with the dynamical symmetry. 
In Sec.~\ref{sec:derivation}, 
mean fields of $\mu$-SB are derived through minimization of energy functionals 
constructed from the underlying Lie algebra.
In Sec.~\ref{sec:block-diagonalization}, 
we show that the quadratic part of an effective Lagrangian of Nambu-Goldstone modes 
can be block-diagonalized for $\mu$-symmetry breaking. 
In Sec.~\ref{sec:calculation}, 
we show how to systematically calculate homotopy groups of topological excitations 
for $\mu$-symmetry breaking. 
In Sec.~\ref{sec:sun}, 
we apply our method to a $U(N)$-symmetric system. 
In Sec.~\ref{sec:application}, 
the cases of higher-dimensional representations are discussed from the standpoint of $\mu$-symmetry breaking 
by using examples of spin-2 Bose-Einstein condensates (BECs) \cite{Koashi00, Ciobanu00, Ueda02} 
and spin-1 color superconductors \cite{Schmitt05, Brauner08}. 
In Sec.~\ref{sec:conclusion}, 
we conclude this paper. 
In Appendix \ref{sec:D}, 
we prove some formulas on homotopy groups used in Sec.~\ref{sec:calculation}.

\section{\label{sec:definition}$\mu$-symmetry breaking}

We consider a quantum field theory whose symmetry group $G$ 
is described by a finite-dimensional unitary representation $\bm{R}$: 
\begin{eqnarray}
\phi_i \mapsto \sum_j \left[\exp\left(i  \sum_{a=1}^d T_a t_a \right) \right]_{ij} \phi_j
, 
\end{eqnarray}
where $\{\phi_i\}_i$ is a set of fields of particles, $T_a$ is an element of the Lie algebra $\mathfrak{g} = \{T_a \}_{a=1}^d$ 
constituted from finite-dimensional Hermitian matrices of the representation $\bm{R}$, respectively. 
Here, $d$ is the dimension of $G$ and 
$t_a$'s are real parameters. 
We denote the Noether charge associated with the generator $T_a$ by $\hat{Q}_{T_a}$. 
For the present discussion, we do not need to specify quantum statistics of particles and the system can be defined either on a lattice or in continuous space.
We assume that the Lie algebra $\mathfrak{g}$ of the symmetry group is the direct product of a simple compact Lie algebra $\bar{\mathfrak{g}}$ and $\mathfrak{u}(1)=\{x I| x \in \mathbb{R}\}$: 
\begin{eqnarray}
\mathfrak{g}=\bar{\mathfrak{g}} \oplus \mathfrak{u}(1)
, 
\end{eqnarray}
where $I$ is the identity matrix. 
The particle-number operator is the Noether charge associated with the generator $I$. 

The key ingredient in the following analysis is the quadratic Casimir invariant defined by 
\begin{eqnarray}
C_2^{\bar{\Gg}}(\bm{v}) := \sum_{a=1}^{\bar{d}} \left( \bm{v} | T_a | \bm{v} \right)^2 
, 
\end{eqnarray}
where $\left| \bm{v} \right)$ and $\bar{d}=d-1$ are a vector in the representation $\bm{R}$ and the dimension of $\bar{\Gg}$, respectively. 
Let $\tbr{H_b}_{b=1}^{\bar{r}}$ be the Cartan subalgebra of $\bar{\Gg}$, i.e. the maximal commutative subalgebra of $\bar{\Gg}$, 
where $\bar{r} = \mathrm{rank} \ \bar{\mathfrak{g}}$ is the rank of the Cartan subalgebra. 
Let $| \bm{\mu} )$ be a weight vector which is a simultaneous eigenstate of $\tbr{H_b}_{b=1}^{\bar{r}}$: 
\begin{eqnarray}
H_b | \bm{\mu} ) &=& \mu_b | \bm{\mu} ) \ (b= 1,2,\cdots,\bar{r})
, \label{eq:eigenweight}\\
\bm{\mu} &=& {}^t(\mu_1, \mu_2, \cdots, \mu_{\bar{r}})
, 
\end{eqnarray}
where $t$ denotes the transpose. 
The highest weight $\hmustate$ is the weight vector 
that maximizes the expectation value of the quadratic Casimir invariant $C_2^{\bar{\Gg}}$: 
\begin{eqnarray}
C_2^{\bar{\Gg}}(\hmu) &=& \sum_{a=1}^{\bar{d}} \left( \hmu | T_a | \hmu \right)^2 
\nonumber \\
&=& \max_{\phi, \left( \phi | \phi \right) = 1} \sum_{a=1}^{\bar{d}} \left( \phi | T_a | \phi \right)^2. 
\end{eqnarray}
The zero weight $\zmustate$ is the weight vector that has the zero eigenvalue of Eq.~(\ref{eq:eigenweight}) 
and therefore minimizes the expectation value of $C_2^{\bar{\Gg}}$ to $0$: 
\begin{eqnarray}
C_2^{\bar{\Gg}}(\zmu) = 0. 
\end{eqnarray}
An irreducible representation of $\bar{\Gg}$ is uniquely determined by its highest weight $\hmu$ \cite{Georgi99}, 
and we denote a complete set of weight vectors belonging to $\hmu$ as $W[\hmu]$. 

We now introduce the concept of $\mu$-symmetry breaking. 
Let $\vev$ be an order parameter. 
If $\vev$ transforms in a low-dimensional representation of the symmetry group $G$ 
in the presence of off-diagonal long-range order (ODLRO), 
$\vev$ will be shown to take either of the following forms:
\begin{eqnarray}
\vev &=& \hmustate
, \label{eq.musb1}\\
\vev &=& \zmustate
. \label{eq.musb2} 
\end{eqnarray}
If the lattice of space, which we denote by $L$, is free from frustration in the presence of diagonal long-range order (DLRO), 
the mean-field ground state $| GS \rangle$ will be shown to take either of the following forms:
\begin{eqnarray}
| GS \rangle &=& \bigotimes_{i \in L} \hmustate_i 
, \label{eq.musb3}\\
| GS \rangle &=& \bigotimes_{u \in \mathcal{U}} \bigotimes_{i \in u} \left| \bm{\mu}_i \right)_i
, \label{eq.musb4} 
\end{eqnarray} 
where $\left| \bm{\mu}_i \right)_i$, $u$, and $\mathcal{U}$ denote 
a simultaneous eigenstate of $\tbr{H_b}_{b=1}^{\bar{r}}$ at lattice site $i$, 
a unit cell of an ordered state, 
and the lattice constituted from the entire set of the unit cells, 
respectively. 
In Eq.~(\ref{eq.musb4}), 
the set $\{ \bm{\mu}_i \}_i$ is chosen so that 
the expectation value of $\widehat{C}_2^{\Gg}$ within each unit cell vanishes 
(see Eq.~(\ref{eq.casimirunit}) and the following explanation for detail): 
\begin{eqnarray}
\langle \widehat{C}_2^{\Gg} \rangle = 
\sum_{a=1}^{\bar{d}} \left[
\sum_{i \in u} {}_i( \bm{\mu}_i | T_a | \bm{\mu}_i )_i 
\right]^2 = 0
. \label{eq.musb5} 
\end{eqnarray}
As is shown in Eq.~(\ref{eq.casimirunit}), 
Eq.~(\ref{eq.musb5}) implies that the sum of the weight vectors within each unit cell vanishes. 

The derivations of Eqs.~(\ref{eq.musb1}) - (\ref{eq.musb4}) will be shown in the following section. 
We call these four types of symmetry breaking $\mu$-symmetry breaking ($\mu$-SB), 
which is characterized by the combination of the highest or zero weight and ODLRO or DLRO. 
Prototypical examples of $\mu$-SB include 
the ferromagnetic phase and 
the polar (antiferromagnetic) phase of a spin-1 BEC \cite{Ohmi98, Ho98}, 
classical ferromagnets (FMs), 
and classical antiferromagnets (AFMs) (see Table \ref{tab:table2}). 

The order parameter of a spin-1 BEC is described by a three-dimensional complex vector 
\begin{eqnarray}
\langle\bm{\phi}\rangle = 
\left( 
\begin{array}{c}
\langle\phi_1\rangle \\
\langle\phi_0\rangle \\
\langle\phi_{-1}\rangle
\end{array}
\right)
, 
\end{eqnarray}
and the symmetry group $G$ is $U(1) \times SO(3)$. 
The Cartan generator of $SO(3)$ is the $S_z$-operator defined by 
\begin{eqnarray}
S_z = \left( \begin{array}{ccc} 1&0&0 \\ 0&0&0 \\ 0&0&-1 \end{array} \right)
. 
\end{eqnarray}
The quadratic Casimir invariant of this system is 
\begin{eqnarray}
C_2^{\mathfrak{so}(3)} = \bm{S} \cdot \bm{S}
, 
\end{eqnarray}
where $\bm{S} = (S_x, S_y, S_z)$ is the vector of spin operators in the Cartesian representation. 
In the ferromagnetic phase, the order parameter has the form 
\begin{eqnarray}
\langle\bm{\phi}\rangle_{\mr{FM}} = \left( \begin{array}{c}1\\0\\0 \end{array} \right)
.  
\end{eqnarray}
This is the eigenstate of $S_z$ with eigenvalue $1$ and maximizes the expectation value of 
$\widehat{C}_2^{\mathfrak{so}(3)} = \widehat{\bm{S}} \cdot \widehat{\bm{S}}$: 
\begin{eqnarray}
\langle \widehat{C}_2^{\mathfrak{so}(3)} \rangle = 
\langle \widehat{\bm{S}} \rangle \cdot \langle \widehat{\bm{S}} \rangle = 1
.  
\end{eqnarray}
Thus, the ferromagnetic phase of a spin-1 BEC is characterized by 
$\mu$-SB with ODLRO and the highest weight $\hmu$. 
In the polar phase, the order parameter has the form 
\begin{eqnarray}
\langle\bm{\phi}\rangle_{\mr{polar}} = \left( \begin{array}{c}0\\1\\0 \end{array} \right)
.  
\end{eqnarray}
This is the eigenstate of $S_z$ with eigenvalue $0$ and minimizes the expectation value of 
$\widehat{C}_2^{\mathfrak{so}(3)}$: 
\begin{eqnarray}
\langle \widehat{C}_2^{\mathfrak{so}(3)} \rangle = 0
.  
\end{eqnarray}
Thus, the polar phase of a spin-1 BEC is characterized by 
$\mu$-SB with ODLRO and the zero weight $\zmu$. 

For both the classical FM and the classical AFM, 
the quadratic Casimir invariant is again the square of a spin $\bm{S}$: 
\begin{eqnarray}
C_2^{\mathfrak{so}(3)} = \bm{S} \cdot \bm{S}
. 
\end{eqnarray}
Let $S$ be the spin quantum number. 
The highest-weight state $|\hmu)$ corresponds to the eigenstate of $S_z$ with the highest magnetic quantum number $m_z$, i.e., $|\hmu) = |m_z = S)$. 
The mean-field ground state $| GS \rangle_{\mr{FM}}$ of the classical FM 
represents a uniform alignment of the highest magnetic quantum-number state $|m_z = S)$ on every site of the lattice $L$: 
\begin{eqnarray}
| GS \rangle_{\mr{FM}} &=& \bigotimes_{i \in L} |m_z = S)_i
. 
\end{eqnarray}
Therefore, the classical FM is characterized by $\mu$-SB with DLRO and $\hmu$. 
On the other hand, 
the mean-field ground state $| GS \rangle_{\mr{AFM}}$ of the classical AFM 
is described by a uniform alignment of the highest magnetic quantum-number state $|m_z = S)$ on the sites of one sublattice $L_A$ 
and that of the lowest magnetic quantum-number state $|m_z = -S)$ on the other sublattice $L_B$: 
\begin{eqnarray}
| GS \rangle_{\mr{AFM}} &=& \bigotimes_{u \in \mathcal{U}} \left( |m_z = S)_{u_A} \otimes |m_z = -S)_{u_B} \right)
, 
\end{eqnarray}
where $u_A$ and $u_B$ indicate the sites in each unit cell $u$ belonging to $L_A$ and $L_B$, respectively. 
Within each unit cell, the total magnetization vanishes. 
Therefore, this phase is characterized by $\mu$-SB with DLRO and $\zmu$.

\begin{table}
\caption{\label{tab:table2}
Examples of four types of $\mu$-symmetry broken phases. 
Here, $\hmu$ and $\zmu$ represent the highest-weight and zero-weight vectors, respectively. 
The order parameters for ODLRO are given by Eqs.~(\ref{eq.musb1}) and (\ref{eq.musb2}) 
and the ground states for DLRO are given by Eqs.~(\ref{eq.musb3}) and (\ref{eq.musb4}). 
}
\begin{ruledtabular}
\begin{tabular}{l|ll}
 & ODLRO & DLRO \\
\hline
$\hmu$ & ferromagnetic spin-1 BEC & ferromagnet \\
$\zmu$ & polar spin-1 BEC & antiferromagnet 
\end{tabular}
\end{ruledtabular}
\end{table}

\section{\label{sec:derivation} Derivations of mean fields of $\mu$-SB}
In this section, 
we show that mean fields described by the highest or zero-weight states are obtained
through the minimization of an energy functional constructed from Casimir invariants of the underlying Lie algebra $\bar{\Gg}$. 

We first discuss the case of ODLRO. 
In this case, the energy functional can be obtained in a manner similar to the case of the dynamical symmetry \cite{Iachello06, Iachello84, Uchino08}. 
However, as shown later, for $\mu$-SB we can block-diagonalize the quadratic part of an effective Lagrangian of NG modes and 
systematically calculate homotopy groups of topological excitations, 
neither of which can be done from the dynamical symmetry alone. 
For the case of the lowest-dimensional representation such as a BEC in degenerate $N$-component bosons 
the mean-field energy functional $V(\vev)$ can be expressed up to the fourth order in $\vev$ as
\begin{eqnarray}
V(\vev)= - c|\vev|^2+c_0 |\vev|^4
, \label{eq.efunc1}
\end{eqnarray}
where $c$ and $c_0$ are real constants. 
Since any minimizer $\vev$ of $V(\vev)$ in the lowest-dimensional representation can be transformed into $\hmustate$ by an appropriate element of $G$, 
the energy functional is minimized for 
\begin{eqnarray}
\vev = \hmustate
. 
\end{eqnarray} 

For the case of the next lowest-dimensional representation such as a spin-1 BEC \cite{Ohmi98, Ho98} and an $s$-wave superfluid in degenerate $N$-component fermions \cite{Modawi97, Cherng07, Rapp07, Cazalilla14} $V(\vev)$ can be expressed in terms of $|\vev|^2$ and the quadratic Casimir invariant as
\begin{eqnarray}
V(\vev)= - c|\vev|^2+ c_0 |\vev|^4+ c_1 C_2^{\bar{\Gg}} (\vev)
, \label{eq.efunc2}
\end{eqnarray}
where $c,c_0$, and $c_1$ are real constants. 
For the case of a ferromagnetic interaction between condensate particles with $c_1<0$, 
the energy functional is minimized for  
\begin{eqnarray}
\vev = \hmustate
. 
\end{eqnarray} 
For the case of an antiferromagnetic interaction with $c_1>0$, 
the energy functional is minimized for 
\begin{eqnarray}
\vev = \zmustate
, 
\end{eqnarray} 
if the representation includes the zero-weight state $\zmustate$. 
This condition is satisfied for low-dimensional Lie algebras such as $\mathfrak{su}(2)$ and $\mathfrak{so}(3)$. 
The general case of $\bar{\Gg} = \mathfrak{su}(N)$ will be discussed in Sec.~\ref{sec:sun}. 
Since the energy functional can be written in neither form of Eq.~(\ref{eq.efunc1}) nor Eq.~(\ref{eq.efunc2}) in higher-dimensional representations, 
the ground states are no longer described by $\mu$-SB. 
Such a case will be discussed in Sec.~\ref{sec:application}.

\begin{figure}
\begin{center}
\includegraphics[width=7.5cm]{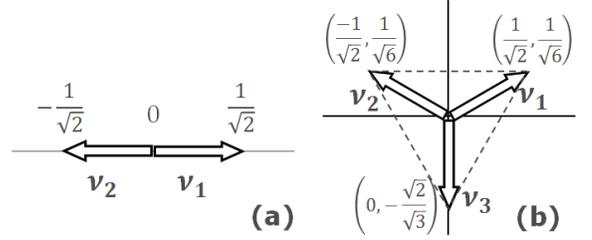}
\caption{\label{fig:weight} 
(a) Weight vectors of the two-dimensional representation of the $\mathfrak{su}(2)$-Lie algebra. 
The weight vectors are two one-dimensional vectors $\bm{\nu}_1=1/\sqrt{2}$ and $\bm{\nu}_2=-1/\sqrt{2}$, 
which have opposite directions with the same magnitude $1/\sqrt{2}$. 
In this representation, the pair of the weight vectors that satisfy Eq.~(\ref{eq.DLROmuz1}) 
is uniquely determined to be $(\bm{\nu}_1, \bm{\nu}_2)$. 
(b) Weight vectors of the three-dimensional representation of the $\mathfrak{su}(3)$-Lie algebra. 
The weight vectors are three two-dimensional vectors $\bm{\nu}_1=(1/\sqrt{2}, 1/\sqrt{6}), \bm{\nu}_2=(-1/\sqrt{2}, 1/\sqrt{6})$, and $\bm{\nu}_3=(0, -\sqrt{2/3})$ 
from the origin to each of the three apexes of the equilateral triangle. 
The sum of these three weight vectors vanishes. 
Three pairs of the weight vectors, 
$(\bm{\nu}_1, \bm{\nu}_2)$, $(\bm{\nu}_2, \bm{\nu}_3)$, and $(\bm{\nu}_3, \bm{\nu}_1)$, 
satisfy Eq.~(\ref{eq.DLROmuz1}). 
In both (a) and (b), 
the weight vectors are normalized so as to satisfy Eq.~(\ref{eq:sunweightvectors}). 
}
\end{center}
\end{figure}

We next discuss the case of DLRO. 
We assume that particles are placed on a lattice $L$ whose geometry is free from frustration. 
A prototypical example of DLRO is described by the Heisenberg-type Hamiltonian 
\begin{eqnarray}
H = -J \sum_{\langle i,j\rangle} \sum_{a=1}^{\bar{d}} T_{a, i} T_{a, j}
, 
\end{eqnarray}
where $\langle i,j \rangle$ represents a pair of nearest-neighbor sites $i$ and $j$, 
and $\{T_{a, i} \}_{a=1}^{\bar{d}}$ is a set of generators of a simple compact Lie algebra $\bar{\Gg}$ on site $i$. 
For the case of a ferromagnetic interaction with $J>0$, 
the ground state $| GS \rangle$ is written as 
\begin{eqnarray}
| GS \rangle = \bigotimes_{i \in L} \hmustate_i
, 
\end{eqnarray}
which coincides with Eq.~(\ref{eq.musb3}). 
For the case of an antiferromagnetic interaction with $J<0$, 
it follows from the frustration-free assumption 
that the mean-field classical ground state $| GS \rangle$ is obtained by a tensor product of the state on site $i$ 
that minimizes the interaction energy with its neighboring sites \cite{Lacroix11}. 
Therefore, the mean-field ground state can be written in terms of a site-factorized wave function as 
\begin{eqnarray}
| GS \rangle = \bigotimes_{u \in \mathcal{U}} \bigotimes_{i \in u} \left| \bm{\mu}_i \right)_i
, \label{eq.sitefactrized}
\end{eqnarray}
where $\left| \bm{\mu}_i \right)_i$, $u$, and $\mathcal{U}$ denote 
a simultaneous eigenstate of $\tbr{H_b}_{b=1}^{\bar{r}}$ at lattice site $i$, 
a unit cell of an ordered state, 
and the lattice constituted from the entire set of the unit cells, 
respectively. 
In Eq.~(\ref{eq.sitefactrized}), 
two weight vectors $\bm{\mu}_i$ and $\bm{\mu}_j$ at nearest-neighbor sites $i$ and $j$ should satisfy
\begin{eqnarray}
\left( \bm{\mu}_i | \bm{\mu}_j \right) = \min\tbr{ \left(\bm{\mu}_k|
\bm{\mu}_l\right) | \bm{\mu}_k, \bm{\mu}_l \in W[\hmu]}
\label{eq.DLROmuz1}
\end{eqnarray}
to minimize the nearest-neighbor interaction energy $\sum_{a=1}^{\bar{d}} T_{a, i} T_{a, j}$. 
We note that we do not impose Eq.~(\ref{eq.musb5}) on Eq.~(\ref{eq.sitefactrized}) so far. 
Since the expectation value $( \bm{\mu}_i \left| T_a \right| \bm{\mu}_i )$ vanishes for any off-diagonal matrix $T_a$, 
which is nothing but the raising or lowering operator of the Cartan canonical form (see Sec.~\ref{sec:block-diagonalizationA}), 
the expectation value of $\widehat{C}_2^{\bar{\Gg}}$ within each unit cell $u$ 
can be calculated from Eqs.~(\ref{eq:eigenweight}) and (\ref{eq.sitefactrized}) 
by considering the contribution from the Cartan generators $\{ H_b \}_{b=1}^{\bar{r}}$ alone. 
We thus obtain 
\begin{eqnarray}
\langle \widehat{C}_2^{\bar{\Gg}} \rangle 
&=& 
 \sum_{b=1}^{\bar{r}} \left[ 
\sum_{i \in u} {}_i( \bm{\mu}_i \left| H_b \right| \bm{\mu}_i )_i
\right]^2
\nonumber\\
&=& 
 \sum_{b=1}^{\bar{r}} \left[ 
\sum_{i \in u} (\bm{\mu}_i)_b
\right]^2
\nonumber\\
&=&  \left\lVert \sum_{i \in u} \bm{\mu}_i \right\rVert^2
, \label{eq.casimirunit}
\end{eqnarray}
where $\lVert \bm{x} \rVert$ denotes the magnitude of the vector $\bm{x}$. 
If there exist a pair of weight vectors with opposite directions such as the case of $SU(2)$ (see Fig. \ref{fig:weight} (a)), 
Eq.~(\ref{eq.DLROmuz1}) is satisfied 
and the sum of the weight vector in Eq.~(\ref{eq.casimirunit}) vanishes within each unit cell. 
Thus, the ground-state wave function satisfies Eq.~(\ref{eq.musb5}). 
However, 
in a larger group such as $SU(3)$, 
it is known that there do not exist two weight vectors with opposite directions 
and that there is more than one pair of weight vectors that satisfy Eq.~(\ref{eq.DLROmuz1}) (see Fig. \ref{fig:weight} (b)) \cite{Georgi99}. 
For such cases, the mean-field ground states represented by Eq.~(\ref{eq.sitefactrized}) are degenerate \cite{Papanicolaou88}. 
Therefore, we have to consider higher-order contributions arising from quantum fluctuations to determine the ground state. 
This can be done by using the flavor-wave theory \cite{Papanicolaou88, Joshi99, Corboz11, Corboz12}. 
The Hamiltonian $H_{\mr{fw}}$ of quantum fluctuations is given as follows \cite{Corboz11, Corboz12}:
\begin{eqnarray}
H_{\mr{fw}} &=& \sum_{\langle i, j \rangle} A_{ij}^\dagger A_{ij}
, 
\end{eqnarray}
where
\begin{eqnarray}
A_{ij} &=& b_{\bm{\mu}_i, j} + b_{\bm{\mu}_j, i}^\dagger
, 
\end{eqnarray}
and $b_{\bm{\mu}_i, j}$ is the annihilation operator of a boson with with weight vector $\bm{\mu}_i$ at site $j$. 
The expectation value of $H_{\mr{fw}}$ is minimized when 
\begin{eqnarray}
\langle GS |  A_{ij}^\dagger A_{ij} | GS \rangle = 0 \Leftrightarrow 
A_{ij} | GS \rangle = 0 
 \ \mr{for} \ \forall i, j
. 
\label{eq.FW}
\end{eqnarray}
Let $i,j$ and $j,k$ be pairs of nearest-neighbor sites. 
Then, $i$ and $k$ have the common nearest-neighbor site $j$. 
It follows from Eq.~(\ref{eq.FW}) that 
\begin{eqnarray}
0 &=& [A_{ij}, A_{jk}] | GS \rangle = [b_{\bm{\mu}_i, j}, b_{\bm{\mu}_k, j}^\dagger] | GS \rangle 
\nonumber\\
&=& \delta_{\bm{\mu}_i, \bm{\mu}_k} | GS \rangle
, 
\end{eqnarray}
and hence we have 
\begin{eqnarray}
\bm{\mu}_i &\ne& \bm{\mu}_k  
\label{eq.DLROmuz2}
\end{eqnarray}
for any pair of sites $i$ and $k$ that have a common nearest-neighbor site $j$. 
Thus, the ground state must satisfy both Eqs.~(\ref{eq.DLROmuz1}) and (\ref{eq.DLROmuz2}). 
Equation (\ref{eq.DLROmuz1}) implies that 
two weight vectors, $\bm{\mu}_i$ and $\bm{\mu}_j$, must give the minimum value of the inner product $(\bm{\mu}_i| \bm{\mu}_j)$ 
for any pair of nearest-neighbor sites $i$ and $j$. 
Equation (\ref{eq.DLROmuz2}) implies that 
two weight vectors, $\bm{\mu}_i$ and $\bm{\mu}_k$, must be different 
for any pair of sites $i$ and $k$ that share a common nearest-neighbor site. 
As a consequence, 
the sum of the weight vectors within each unit cell in Eq.~(\ref{eq.casimirunit}) tend to vanish 
and an ordered state with the vanishing expectation value of $\widehat{C}_2^{\bar{\Gg}}$ within each unit cell is favored. 
Thus, the ordered state is characterized by $\mu$-SB with $\zmu$.

\section{\label{sec:block-diagonalization}Effective Lagrangian of NG modes for $\mu$-SB}

\subsection{\label{sec:block-diagonalizationA}Cartan canonical form and the generalized magnetization}
The analyses of NG modes and topological excitations in $\mu$-SB can be done conveniently 
in terms of a special basis of the Lie algebra $\bar{\mathfrak{g}}$ known as the Cartan canonical form \cite{Georgi99}:
\begin{eqnarray}
\bar{\mathfrak{g}} = \{ \{ H_b \}_{b=1}^{\bar{r}}, \{E_{\Bal}^R, E_{\Bal}^I \}_{\Bal \in R_+} \}
, \label{eq:Cartanform}
\end{eqnarray}
where $R$ and $I$ indicate the real and imaginary parts of the raising operators $E_{\pm \Bal}$ of the Cartan canonical form: 
\begin{eqnarray}
E_{\bm{\alpha}}^R := \frac{E_{\bm{\alpha}} + E_{-\bm{\alpha}}}{\sqrt{2}}
, \ \ 
E_{\bm{\alpha}}^I := \frac{E_{\bm{\alpha}} - E_{-\bm{\alpha}}}{\sqrt{2} i}
.
\end{eqnarray}
The Cartan canonical form (\ref{eq:Cartanform}) is a generalization of the basis of the $\mathfrak{su}(2)$ Lie algebra $\{S^z, \{S^x, S^y\}\}$, 
which decomposes the generators into the diagonal matrices $\{ H_b \}_{b=1}^{\bar{r}}$ 
and the off-diagonal ones $\{E_{\Bal}^R, E_{\Bal}^I \}_{\Bal \in R_+}$, 
where $\Bal$ is an $\bar{r}$-dimensional real vector known as a positive root 
and $R_+$ denotes the set of all the positive roots. 
Here the positive roots distinguish different $\mathfrak{su}(2)$-subalgebras of $\bar{\mathfrak{g}}$. 
Defining $H_{\Bal}$ for $\Bal \in \mathbb{R}^{\bar{r}}$ by 
\begin{eqnarray}
H_{\Bal} &:=& \sum_{b=1}^{\bar{r}} \alpha_b H_b
,
\end{eqnarray}
where
\begin{eqnarray}
\Bal &=& {}^t(\alpha_1, \alpha_2, \cdots, \alpha_{\bar{r}})
, 
\end{eqnarray}
we obtain a triad $\bm{S}_{\Bal}$ for a positive root $\Bal$ as 
\begin{eqnarray}
\bm{S}_{\Bal} = (E_{\Bal}^R, E_{\Bal}^I, H_{\Bal})
. \label{eq:generalizedmagnetization}
\end{eqnarray} 
The triad $\bm{S}_{\Bal}$ forms an $\mathfrak{su}(2)$ subalgebra of $\bar{\Gg}$ analogous to the $\mathfrak{su}(2)$ spin algebra. 
In fact, it satisfies the following commutation relations: 
\begin{eqnarray}
&&[E_{\Bal}^R, E_{\Bal}^I] = i (\Bal, \Bal) H_{\Bal}
 \ \mr{for} \ \forall \Bal \in R_+
, \label{eq:Halpha1}\\
&&[E_{\Bal}^I, H_{\Bal}] =  i  (\Bal, \Bal) E_{\Bal}^R
 \ \mr{for} \ \forall \Bal \in R_+
, \label{eq:Halpha2}\\
&&[H_{\Bal}, E_{\Bal}^R] = i  (\Bal, \Bal) E_{\Bal}^I
 \ \mr{for} \ \forall \Bal \in R_+
, \label{eq:Halpha3}\\
&&[H_{\Bal}, H_{\Bal^\prime}] = 0
 \ \mr{for} \ \forall \Bal, \forall \Bal^\prime \in R_+
. \label{eq:Halpha4}
\end{eqnarray}
We refer to $\bm{S}_{\Bal}$ as a generalized magnetization in analogy with the $\mathfrak{su}(2)$ spin $\bm{S}$. 
As shown later, the textures of NG modes and topological excitations 
are described in terms of the generalized magnetization $\bm{S}_{\Bal}$. 

\subsection{Broken generators for $\mu$-SB}
To determine the quadratic part of an effective Lagrangian of NG modes, 
we first prove the following theorem on the spaces of broken generators $\Gg/\Gh$, 
where $\Gh$ is the Lie algebra of the remaining (unbroken) symmetry of the state. 

\begin{theo}
\label{theo:unbroken}
Consider $\mu$-SB phases and define the sets of positive root vectors $R_H$, $R_0^{\mr{OD}}$, and $R_0^{\mr{D}}$ by 
\begin{eqnarray}
R_H &:=& \{\Bal \in R_+| (\Bal, \hmu) \ne 0 \}
, \label{eq.RH}\\
R_0^{\mr{D}} &:=& \bigcup_{i \in u} \{\Bal \in R_+| \left| \bm{\mu}_i - \Bal \right) \in W[\hmu] \}
, \label{eq.R0D}\\
R_0^{\mr{OD}} &:=& \{\Bal \in R_+| \left| \zmu - \Bal \right) \in W[\hmu] \}
. \label{eq.R0OD}
\end{eqnarray}
Then the bases of the space of broken generators $\Gg/\Gh$ are given by 
\begin{eqnarray}
&&\mathrm{basis}(\Gg/\Gh)
\nonumber\\
 &=& 
\begin{cases}
\tbr{ E_{\Bal}^R, E_{\Bal}^I |\Bal \in R_H }
&\ \mr{for} \ \mr{DLRO \ and \ } \hmu; \\
\tbr{ E_{\Bal}^R, E_{\Bal}^I |\Bal \in R_0^{\mr{D}} }
&\ \mr{for} \ \mr{DLRO \ and \ } \zmu; \\ 
\tbr{ E_{\Bal}^R, E_{\Bal}^I |\Bal \in R_H } \cup \tbr{I}
&\ \mr{for} \ \mr{ODLRO \ and \ } \hmu; \\
\tbr{ E_{\Bal}^R, E_{\Bal}^I |\Bal \in R_0^{\mr{OD}} } \cup \tbr{I} 
&\ \mr{for} \ \mr{ODLRO \ and \ } \zmu. 
\end{cases}
\nonumber \\
\label{eq.brokengenerator}
\end{eqnarray}
\end{theo}
\textit{Proof}:

First, we consider the DLRO with $\hmu$. 
From the mean-field ground state in Eq.~(\ref{eq.musb3}), 
unitary transformations generated by $E_{\Bal}^{R,I}$ leave the ground state unchanged up to a global phase if and only if 
\begin{eqnarray}
&&E_{\Bal}^{R,I} \left| \hmu \right) = 0
\nonumber\\
&\Leftrightarrow& E_{+\Bal} \left| \hmu \right) = E_{-\Bal} \left| \hmu \right) = 0
. \label{eq.weight11}
\end{eqnarray}
Since $\hmu$ is the highest weight, $E_{+\Bal} \left| \hmu \right)$ always vanishes \cite{Georgi99}, 
so that Eq.~(\ref{eq.weight11}) is equivalent to 
\begin{eqnarray}
E_{-\Bal} \left| \hmu \right) = 0
\Leftrightarrow
\left| \hmu - \Bal \right) \notin W[\hmu]
\label{eq.weight1}
, 
\end{eqnarray}
which, in turn, is equivalent to 
\begin{eqnarray}
\Bal \notin R_H
\label{eq.weight2}
. 
\end{eqnarray}
This equivalence can be shown as follows. 
From Eq.~(\ref{eq.weight1}), 
we obtain 
\begin{eqnarray}
0 &=& \left[E_{\Bal}^R, E_{\Bal}^I \right] \left| \hmu \right) = (\Bal, \Bal) H_{\Bal} \left| \hmu \right) 
\nonumber\\
&=& (\Bal, \Bal) (\hmu, \Bal) \left| \hmu \right)
. 
\end{eqnarray}
Since $(\Bal, \Bal) \ne 0$ and $\left| \hmu \right) \ne 0$, we obtain $(\hmu, \Bal) = 0$ 
and hence Eq.~(\ref{eq.weight2}) from the definition of $R_H$. 
Conversely, we can derive Eq.~(\ref{eq.weight1}) by assuming Eq.~(\ref{eq.weight2}). 
The Weyl reflection \cite{Georgi99} of $\hmu + \Bal$ with respect to $\Bal$ is 
\begin{eqnarray}
(\hmu + \Bal) -2 \frac{(\hmu + \Bal, \Bal)}{(\Bal, \Bal)} \Bal
= \hmu - \Bal
.
\end{eqnarray}
Since $\left| \hmu \right)$ is the highest weight, 
$\left| \hmu + \Bal \right)$ is not a weight vector, nor is $\left| \hmu - \Bal \right)$. 
Thus, Eq.~(\ref{eq.weight1}) is obtained. 
Unitary transformations generated by $\tbr{ H_b }_{b=1}^{\bar{r}}$ and $I$ change the ground state $| GS \rangle$ only by a global phase factor. 
Therefore, these generators are not broken ones, 
which completes the proof of the first row of Eq.~(\ref{eq.brokengenerator}). 

Second, we consider the DLRO with $\zmu$. 
From the mean-field ground state in Eq.~(\ref{eq.musb4}), 
unitary transformations generated by $E_{\Bal}^{R,I}$ leave $|GS\rangle$ invariant if and only if 
\begin{eqnarray}
E_{\Bal}^{R,I} \left| \bm{\mu}_i \right) = 0
 \ \mr{for} \ \forall i \in u
. 
\end{eqnarray}
This condition is equivalent to $\Bal \notin R_0^{\mr{OD}}$. 
From the discussions similar to the case of DLRO and $\hmu$, 
all of the Cartan generators are not broken ones, 
which completes the proof of the second row of Eq.~(\ref{eq.brokengenerator}). 

Third, we consider the ODLRO with the highest weight, i.e. $\vev = \hmustate$. 
Unitary transformations generated by $E_{\Bal}^{R,I}$ leave $|GS\rangle$ invariant 
if and only if Eq.~(\ref{eq.weight1}) is satisfied. 
From above discussion, this condition is equivalent to $\Bal \notin R_H$. 
Unitary transformations generated by $\tbr{ H_b }_{b=1}^{\bar{r}}$ and $I$ change $\hmustate$ only by a phase factor. 
This phase shift can be eliminated by taking a linear combination of $\tbr{H_b }_{b=1}^{\bar{r}}$ and $I$ except for the direction of $I$, 
which completes the proof of the third row of Eq.~(\ref{eq.brokengenerator}). 
The proof of the fourth row can be given similarly by replacing $\hmu$ by $\zmu$ and using Eq.~(\ref{eq.weight1}). 
Thus, the proof of Theorem \ref{theo:unbroken} is completed.

\subsection{Effective Lagrangian of NG modes}
We now derive the quadratic part of an effective Lagrangian of NG modes.
We note that in non-relativistic systems the type-2 NG mode with a quadratic dispersion is allowed, 
in contrast to relativistic systems where only the type-1 NG mode with a linear dispersion is allowed \cite{Goldstone61}. 
Let $N_B := \mathrm{dim}(G/H)$ be the number of the broken generators. 
It has been shown that the numbers of type-1 and type-2 NG modes, $n_1$ and $n_2$, 
can be determined only from the ground state $| GS \rangle$ and the set of the Noether charges 
associated with the broken generators of the symmetry group $\{ \hat{T}_{a^\prime} \}_{a^\prime=1}^{N_B}$ 
as follows 
(the prime indicates that the generators are broken ones) \cite{Watanabe12b, Hidaka13}:
\begin{eqnarray}
\left\{
\begin{array}{l}
n_1 + 2 n_2 = \mathrm{dim}(G/H); \\
n_2 = \frac{1}{2} \mathrm{rank} \rho,
\end{array}
\right. \label{D} \\
\rho_{a^\prime b^\prime} = -i
\langle GS | [\hat{Q}_{T_{a^\prime}}, \hat{Q}_{T_{b^\prime}}] | GS \rangle, \label{F}
\end{eqnarray}
where $\rho = \{ \rho_{a^\prime b^\prime} \}_{a^\prime, b^\prime=1}^{N_B}$ is a Kostant-Kirillov symplectic form (K-K form) \cite{Kirillov62, Guillemin90, Kirillov04} 
which is employed in Ref.~\cite{Watanabe11} in the context of NG modes. 
The basis of the Lie algebra that block-diagonalizes Eq.~(\ref{F}) is constituted from a set of canonically conjugate pairs and generates type-2 NG modes \cite{Watanabe12b, Hidaka13}. 
While the types and the numbers of NG modes can be found from Eqs.~(\ref{D}) and (\ref{F}), 
the dynamics and the corresponding broken generator of the NG mode cannot be determined from them 
since the quadratic part of an effective Lagrangian is not diagonalized in Refs. \cite{Watanabe12b, Hidaka13}.
Here, we show that for $\mu$-SB, 
the K-K form and the quadratic part of an effective Lagrangian in the fields of NG modes 
can be simultaneously block-diagonalized in terms of the Cartan canonical form. 
Moreover, NG modes are classified into three categories according to their dynamics. 
In fact, we can prove the following theorem.

\begin{theo}
\label{theo:NGmode}
Consider a $\mu$-SB in a non-relativistic system. 
Let $\pi_{\Bal}^R, \pi_{\Bal}^I$ and $\pi_I$ be the fields of the NG mode generated by the broken generators $E_{\Bal}^R, E_{\Bal}^I$ and $I$ 
and define $\Pi_{\Bal}$ by $\Pi_{\Bal} := \pi_{\Bal}^R + i \pi_{\Bal}^I$. 
Then, the quadratic parts of effective Lagrangians $\mathcal{L}_{\mr{eff}}$ in $\Pi_{\Bal}$ and $\pi_I$ can be block-diagonalized 
in terms of $\Pi_{\Bal}$ and $\pi_I$ as follows: 
\begin{eqnarray}
\mathcal{L}_{\mr{eff}} &=& 
\begin{cases}
\sum_{\Bal \in R_H} \mathcal{L}_{\mr{pre}}^{\Bal}
&\ \mr{for} \ \mr{DLRO \ and \ } \hmu; \\
\sum_{\Bal \in R_0^{\mr{D}}} \mathcal{L}_{\mr{osc}}^{\Bal}
&\ \mr{for} \ \mr{DLRO \ and \ } \zmu; \\ 
\mathcal{L}_{\mr{pha}} + \sum_{\Bal \in R_H} \mathcal{L}_{\mr{pre}}^{\Bal}
&\ \mr{for} \ \mr{ODLRO \ and \ } \hmu; \\
\mathcal{L}_{\mr{pha}} + \sum_{\Bal \in R_0^{\mr{OD}}} \mathcal{L}_{\mr{osc}}^{\Bal}
&\ \mr{for} \ \mr{ODLRO \ and \ } \zmu, 
\end{cases}
\nonumber\\
\label{effLag}\\
\mathcal{L}_{\mr{pre}}^{\Bal} &=& 
\rho_{\Bal}(\Pi_{\Bal} \partial_t \Pi_{\Bal}^\ast - \Pi_{\Bal}^\ast \partial_t \Pi_{\Bal}) + b_{\Bal} |\nabla \Pi_{\Bal}|^2
, \label{effLagpre}\\
\mathcal{L}_{\mr{osc}}^{\Bal} &=&
b_{\Bal}|\nabla \Pi_{\Bal}|^2 + \bar{b}_{\Bal} |\partial_t \Pi_{\Bal}|^2
, \label{effLagosc}\\
\mathcal{L}_{\mr{pha}} &=& 
g_I (\nabla \pi_I)^2 + \bar{g}_I ( \partial_t \pi_I)^2 
, 
\end{eqnarray}
where $\rho_{\Bal}, b_{\Bal}, \bar{b}_{\Bal}, g_I, \bar{g}_I$ are real constants. 
\end{theo}

\textit{Proof}: 
Consider the quadratic part of an effective Lagrangian in the fields of NG modes in a non-relativistic system. 
Then, the most general form of it can be written as follows \cite{Watanabe12b}: 
\begin{eqnarray}
\mathcal{L}_{\mr{eff}} = 
\sum_{a^\prime, b^\prime=1}^{N_B} & &\left( 
\rho_{a^\prime b^\prime} 
\pi_{a^\prime} \partial_t \pi_{b^\prime} 
 + \frac{1}{2} \bar{g}_{a^\prime b^\prime} \partial_t \pi_{a^\prime} \partial_t \pi_{b^\prime} 
\right.
\nonumber\\
 &&\left. 
+ \frac{1}{2} g_{a^\prime b^\prime} \nabla \pi_{a^\prime} \cdot \nabla \pi_{b^\prime} \right)
,
\label{eq:effective}
\end{eqnarray}
where $\rho$ is the K-K form defined in Eq.~(\ref{F}), 
$\{ \pi_{a^\prime} \}_{a^\prime=1}^{N_B}$ is the set of the fields of NG modes associated with the set of the broken generators $\{ T_{a^\prime} \}_{a^\prime=1}^{N_B}$, 
and $\bar{g}_{a^\prime b^\prime}$ and $g_{a^\prime b^\prime}$ are real constants. 

We first block-diagonalize the quadratic part of an effective Lagrangian. 
Here we consider the case of ODLRO and $\hmu$. 
The proof for the other cases can be made in a similar manner. 
The quadratic forms constructed from $\pi_I, \Pi_{\Bal}^\ast$, and $\Pi_{\Bal}$ are the following six terms: 
\begin{eqnarray}
&&(\pi_I)^2, \pi_I \Pi_{\Bal}^\ast, \pi_I\Pi_{\Bal}; 
\nonumber\\
&&\Pi_{\Bal}\Pi_{\bm{\beta}}, \Pi_{\Bal}^\ast\Pi_{\bm{\beta}}^\ast, \Pi_{\Bal}\Pi_{\bm{\beta}}
. \label{eq:quadraticforms}
\end{eqnarray}
Let $\tbr{H_b^\prime}_{b=1}^{\bar{r}-1}$ be a basis of $(\bar{r}-1)$-dimensional subspace that is orthogonal to $\hmu$. 
Under the unitary transformation generated by 
\begin{eqnarray}
H_{\bm{s}} := \sum_{b=1}^{{\bar{r}}-1} s_a H_b^\prime + \frac{s_{\bar{r}}}{|\hmu|} (H_{\hmu} - I |\hmu|^2) \in \Gh
, 
\end{eqnarray}
$\pi_I$ is invariant since the generator $I$ commutes with $H_{\bm{s}}$. 
Using the commutation relations of the Cartan canonical form 
\begin{eqnarray}
\left[ E_{\Bal}^R, H_{\bm{s}}\right] &=& i (\Bal, \bm{s}) E_{\Bal}^I
, \\
\left[E_{\Bal}^I, H_{\bm{s}} \right] &=& -i (\Bal, \bm{s}) E_{\Bal}^R
, 
\end{eqnarray}
we have
\begin{eqnarray}
\mathrm{e}^{ -i H_{\bm{s}}} (E_{\Bal}^R + i E_{\Bal}^I) \mathrm{e}^{ i H_{\bm{s}}} &=& \mathrm{e}^{ i (\bm{s},\Bal)} (E_{\Bal}^R + i E_{\Bal}^I)
, \\
\mathrm{e}^{ -i H_{\bm{s}}} (E_{\Bal}^R - i E_{\Bal}^I) \mathrm{e}^{ i H_{\bm{s}}} &=& \mathrm{e}^{ - i (\bm{s},\Bal)} (E_{\Bal}^R - i E_{\Bal}^I)
.
\end{eqnarray}
Therefore, the corresponding fields $\Pi_{\Bal}$ and $\Pi_{\Bal}^\ast$ transform under the same unitary transformation into 
$\Pi_{\Bal}\mr{e}^{ i (\bm{s}, \Bal)}$ and $\Pi_{\Bal}^\ast \mr{e}^{ -i (\bm{s}, \Bal)}$, respectively. 
Among the quadratic forms in Eq.~(\ref{eq:quadraticforms}), 
only $(\pi_I)^2$ and $\Pi_{\Bal}^\ast \Pi_{\Bal}$ are invariant 
under the transformations generated by $H_{\bm{s}}$ for any $\bm{s} \in \mathbb{R}^{\bar{r}}$. 
Thus, $\mathcal{L}_{\mr{eff}}$ in Eq.~(\ref{eq:effective}) can be written as 
\begin{eqnarray}
\mathcal{L}_{\mr{eff}} &=& 
\rho_I \pi_I \partial_t \pi_I + g_I (\nabla \pi_I)^2 + \bar{g}_I ( \partial_t \pi_I)^2 
\nonumber\\
&&+ \sum_{\Bal \in R_H}
\left(
\rho_{\Bal}\Pi_{\Bal} \partial_t \Pi_{\Bal}^\ast + \rho_{\Bal}^\prime \Pi_{\Bal}^\ast \partial_t \Pi_{\Bal} \right.
\nonumber\\
&&\left. \ \ \ + b_{\Bal} |\nabla \Pi_{\Bal}|^2 + \bar{b}_{\Bal} |\partial_t \Pi_{\Bal}|^2 
\right)
. 
\end{eqnarray}
where $\rho_{\Bal}, \rho_{\Bal}^\prime, b_{\Bal}$, and $\bar{b}_{\Bal}$ are real constants. 
The first, fourth, and fifth terms correspond to the first term in Eq.~(\ref{eq:effective}) 
and hence to the K-K form in Eq.~(\ref{F}). 
From Eq.~(\ref{F}), 
we have 
\begin{eqnarray}
\rho_I &=& -i \langle  [\hat{Q}_{I}, \hat{Q}_{I}]  \rangle = 0
, \\
\rho_{\Bal} &=& -i \langle  [\hat{Q}_{E_{\Bal}^R} + i \hat{Q}_{E_{\Bal}^I}, \hat{Q}_{E_{\Bal}^R} - i \hat{Q}_{E_{\Bal}^I}]  \rangle 
\nonumber\\
&=& -2 i \langle \hat{Q}_{H_{\Bal}} \rangle 
, \\
\rho_{\Bal}^\prime &=& -i \langle  [\hat{Q}_{E_{\Bal}^R} - i \hat{Q}_{E_{\Bal}^I}, \hat{Q}_{E_{\Bal}^R} + i \hat{Q}_{E_{\Bal}^I}]  \rangle 
\nonumber\\
&=& - \rho_{\Bal}
. 
\end{eqnarray}
Thus, we obtain the block-diagonalized form of the quadratic part of an effective Lagrangian in 
$\Pi_{\Bal}^\ast, \pi_I$, : 
\begin{eqnarray}
\mathcal{L}_{\mr{eff}} &=& g_I (\nabla \pi_I)^2 + \bar{g}_I ( \partial_t \pi_I)^2 
\nonumber\\
&&+ \sum_{\Bal \in R_H}
\left[
\rho_{\Bal} (\Pi_{\Bal} \partial_t \Pi_{\Bal}^\ast -  \Pi_{\Bal}^\ast \partial_t \Pi_{\Bal}) \right.
\nonumber\\
&&\left. \ \ \ + b_{\Bal} |\nabla \Pi_{\Bal}|^2 + \bar{b}_{\Bal} |\partial_t \Pi_{\Bal}|^2
\right]
, \label{eq:effective2}\\
\rho_{\Bal} &=& -2 i \langle \hat{Q}_{H_{\Bal}} \rangle 
. \label{eq:effective1}
\end{eqnarray}
Here, 
$R_H$ represents the set of the positive root $\Bal$ vectors associated with the broken generators. 
For the ODLRO (DLRO) with $\zmu$, 
the quadratic part of an effective Lagrangian is obtained by replacing $R_H$ into $R_0^{\mr{OD}} \ (R_0^{\mr{D}})$. 

Next, we calculate $\langle \hat{Q}_{H_{\Bal}} \rangle$ in Eq.~(\ref{eq:effective1}). 
For $\mu$-SB with $\zmu$, 
$\langle \hat{Q}_{H_{\Bal}} \rangle$ vanishes for any positive root vector $\Bal$ 
because the expectation value of any generator also vanishes. 
Therefore, the term $\rho_{\Bal} (\Pi_{\Bal} \partial_t \Pi_{\Bal}^\ast -  \Pi_{\Bal}^\ast \partial_t \Pi_{\Bal})$ in Eq.~(\ref{eq:effective2}) vanishes 
and hence we obtain the fourth row of Eq.~(\ref{effLag}).  
On the other hand, it follows from the definition of $\mu$-SB for $\hmu$ in Eqs.~(\ref{eq.musb1}) and (\ref{eq.musb2}) that 
\begin{eqnarray}
\langle \hat{Q}_{H_{\Bal}} \rangle &=& N_C (\hmu, \Bal) \ne 0
 \ \mr{for} \ \Bal \in R_H
, 
\end{eqnarray} 
where $N_C$ is the number of particles that contribute to the LRO. 
The term $\bar{b}_{\Bal} |\partial_t \Pi_{\Bal}|^2$ in Eq.~(\ref{eq:effective2}) is much smaller than the term $\rho_{\Bal} (\Pi_{\Bal} \partial_t \Pi_{\Bal}^\ast -  \Pi_{\Bal}^\ast \partial_t \Pi_{\Bal})$ at low energy 
because the former is the second order in $\partial_t$ while the latter is the first order in $\partial_t$. 
Therefore, the term $\bar{b}_{\Bal} |\partial_t \Pi_{\Bal}|^2$ can be ignored and hence we obtain the third row of Eq.~(\ref{effLag}). 
For the case of DLRO, the terms involving $\pi_I$ are absent since $I$ is not a broken generator from Theorem \ref{theo:unbroken}. 
Therefore, we obtain the first and second rows of Eq.~(\ref{effLag}), 
which completes the proof of Theorem \ref{theo:NGmode}. 

\subsection{Three types of NG modes in $\mu$-SB}
From Theorem \ref{theo:NGmode}, 
we find that three types of NG modes arise in $\mu$-SB described by $\pi_I$, $\Pi_{\Bal}$ for $\zmu$, 
and $\Pi_{\Bal}$ for $\hmu$, 
which we call the phason, the oscillaton, and the precesson, respectively (see Table \ref{tab:table3}). 
The dynamics of these three NG modes are similar to those of the phonon in the scalar BEC, the magnon in the AFM, and the magnon in the FM, respectively. 

The phason is the type-1 NG mode which arises from the generator $I$ and 
similar to the phonon in the scalar BEC which describes density fluctuations. 
Although both the oscillaton and the precesson arise from 
the real and imaginary parts, $E_{\Bal}^R$ and $E_{\Bal}^I$, of the raising operators $E_{\pm \Bal}$, 
they are different in their dispersion relation and dynamics. 
These differences arise from the expectation value of the Noether charge $\langle \hat{Q}_{H_{\Bal}} \rangle$ associated with the generator $H_{\Bal}$. 
For $\zmu$, both the expectation value $\langle \hat{Q}_{H_{\Bal}} \rangle$ and the first-order term in $\partial_t$ in the quadratic part of an effective Lagrangian in Eq.~(\ref{eq:effective}) vanish. 
The quadratic part of the effective Lagrangian of the oscillaton associated with the generator $E_{\Bal}^{R,I}$ 
can be given from Eq.~(\ref{effLagosc}) as 
\begin{eqnarray}
\mathcal{L}_{\mr{osc}}^{\Bal} &=& b_{\Bal}|\nabla \Pi_{\Bal}|^2 + \bar{b}_{\Bal} |\partial_t \Pi_{\Bal}|^2
\nonumber\\
&=& b_{\Bal}(\nabla \pi_{\Bal}^R)^2 + \bar{b}_{\Bal} (\partial_t \pi_{\Bal}^R)^2 + b_{\Bal}(\nabla \pi_{\Bal}^I)^2 + \bar{b}_{\Bal} (\partial_t \pi_{\Bal}^I)^2
. 
\nonumber\\
\end{eqnarray}
From the effective Lagrangian, 
two fields $\pi_{\Bal}^R$ and $\pi_{\Bal}^I$ associated with the generators $E_{\Bal}^R$ and $E_{\Bal}^I$ are decoupled, 
producing two independent harmonic oscillations of the generalized magnetization $\bm{S}_{\Bal}$: 
\begin{eqnarray}
\bm{S}_{\Bal}(\bm{x})
&=& \left( 
\begin{array}{c}
\sin[\Delta\sin(\bm{k}\cdot\bm{x})]\cos\phi \\
\sin[\Delta\sin(\bm{k}\cdot\bm{x})]\sin\phi \\
\cos[\Delta\sin(\bm{k}\cdot\bm{x})] \\
\end{array}
\right) 
\nonumber\\
&&\ \mr{for} \ \phi = 0 \ \mathrm{or} \ \frac{\pi}{2}
, 
\label{eq:oscillaton}
\end{eqnarray}
where $\Delta \ (\ll 1)$ represents the amplitude of the oscillatons, 
$\bm{k}$ is the wave number of the oscillatons, 
and $\bm{x}$ is the coordinate in space. 
These modes are reminiscent of magnons in the AFM, 
where two magnons representing harmonic oscillations of the magnetization appear. 
On the other hand, 
for $\hmu$, neither the expectation value $\langle \hat{Q}_{H_{\Bal}} \rangle$ nor 
the first-order term in $\partial_t$ in the quadratic part of the effective Lagrangian in Eq.~(\ref{eq:effective}) vanishes. 
The quadratic part of effective Lagrangian of the precesson associated with the generator $E_{\Bal}^{R,I}$ is given from Eq.~(\ref{effLagpre}) as 
\begin{eqnarray}
\mathcal{L}_{\mr{pre}}^{\Bal} &=& \rho_{\Bal}(\Pi_{\Bal} \partial_t \Pi_{\Bal}^\ast - \Pi_{\Bal}^\ast \partial_t \Pi_{\Bal}) + b_{\Bal} |\nabla \Pi_{\Bal}|^2
\nonumber\\
&=& 2 i \rho_{\Bal}( \pi_{\Bal}^I \partial_t \pi_{\Bal}^R - \pi_{\Bal}^R \partial_t \pi_{\Bal}^I) 
\nonumber\\
&& + b_{\Bal}(\nabla \pi_{\Bal}^R)^2 + b_{\Bal} (\nabla \pi_{\Bal}^I)^2
. 
\end{eqnarray}
From the effective Lagrangian, 
we find that two fields $\pi_{\Bal}^R$ and $\pi_{\Bal}^I$ associated with the generators $E_{\Bal}^R$ and $E_{\Bal}^I$ form a canonical conjugate pair, 
producing a precession mode of the generalized magnetization $\bm{S}_{\Bal}$: 
\begin{eqnarray}
\bm{S}_{\Bal}(\bm{x})
&=& \left( 
\begin{array}{c}
\sin\Delta^\prime \cos(\bm{k}\cdot\bm{x}) \\
\sin\Delta^\prime \sin(\bm{k}\cdot\bm{x}) \\
\cos\Delta^\prime 
\end{array}
\right)
, 
\label{eq:precesson}
\end{eqnarray}
where $\Delta^\prime (\ll 1)$ represents a precession angle. 
This mode is reminiscent of a magnon in the FM, 
where one magnon representing the precession of the magnetization appears.

\begin{table*}
\caption{\label{tab:table3}
Three types of NG modes in $\mu$-SB. 
The second column shows the classification of $\mu$-SB. 
The third, fourth, fifth, and sixth columns show the generators, fields, dynamics, and dispersion relations of the NG modes, respectively, 
where $\pi_I$ is the field of the NG mode generated by the broken generators $I$, 
$\Pi_{\Bal}$ is defined as $\Pi_{\Bal} := \pi_{\Bal}^R + i \pi_{\Bal}^I$ 
with $\pi_{\Bal}^R$ and $\pi_{\Bal}^R$ being the fields of NG modes generated by the broken generators $E_{\Bal}^R$ and $E_{\Bal}^I$, respectively, 
and $\bm{S}_{\Bal}$ is the generalized magnetization defined in Eq.~(\ref{eq:generalizedmagnetization}). 
The last column shows a pair of the NG modes into which each of three types of NG modes, 
$\pi_I, \Pi_{\Bal} (\Bal \in R_0)$ and $\Pi_{\Bal} (\Bal \in R_H)$ decay 
through the interaction Lagrangians in Eqs.~(\ref{eq:intLag1}) and (\ref{eq:intLag2}). 
}
\begin{ruledtabular}
\begin{tabular}{l||l|ll|ll|l}
name & classification & generator & field & dynamics & dispersion & decay processes \\
\colrule
\hline
phason & 
ODLRO with $\hmu$ & $I$ & $\pi_I$ & 
density & linear & 
two phasons $\pi_I$ 
\\
& ODLRO with $\zmu$
&&& fluctuation &&
two oscillatons $\Pi_{\Bal} (\Bal \in R_0)$
\\
&&&&&&
two precessons $\Pi_{\Bal} (\Bal \in R_H)$
\\
\hline
oscillaton & 
DLRO with $\zmu$ & $E_{\Bal}^R$ and $E_{\Bal}^I$ & $\Pi_{\Bal}$ & 
oscillation of & linear &
one phason $\pi_I$ and one oscillaton $\Pi_{\Bal}$ 
\\
& ODLRO with $\zmu$ &
$(\Bal \in R_0)$ & $(\Bal \in R_0)$ & $\bm{S}_{\Bal}$ (Eq.~(\ref{eq:oscillaton})) & &
two oscillatons $\Pi_{\bm{\beta}}$ and $\Pi_{\bm{\gamma}} (\Bal = \bm{\beta} + \bm{\gamma})$ 
\\
\hline
precesson & 
DLRO with $\hmu$& $E_{\Bal}^R$ and $E_{\Bal}^I$ & $\Pi_{\Bal}$ & 
precession of & quadratic & 
one phason $\pi_I$ and one precesson $\Pi_{\Bal}$ 
\\
& ODLRO with $\hmu$ &
$(\Bal \in R_H)$ & $(\Bal \in R_H)$ & $\bm{S}_{\Bal}$ (Eq.~(\ref{eq:precesson})) & &
two precessons $\Pi_{\bm{\beta}}$ and $\Pi_{\bm{\gamma}} (\Bal = \bm{\beta} + \bm{\gamma})$ 
\\
\end{tabular}
\end{ruledtabular}
\end{table*}

\subsection{Differences among three types of NG modes in decay processes}
Although both the phason and the oscillaton have linear dispersions, 
they play distinct roles in decay processes. 
To see this, let us consider $\mu$-SB with ODLRO and $\zmu$. 
Similarly to the proof of Theorem \ref{theo:NGmode}, 
under the unitary transformation generated by $H_{\bm{s}} \in \Gh (\bm{s} \in \mathbb{R}^{\bar{r}})$, 
$\pi_I, \Pi_{\Bal}^\ast$, and $\Pi_{\Bal}$ are transformed into 
$\pi_I, \Pi_{\Bal}^\ast\mr{e}^{ -i (\bm{s}, \Bal)}$, and $\Pi_{\Bal} \mr{e}^{ i (\bm{s}, \Bal)}$, respectively. 
Up to the third order in $\pi_I$, $\pi_{\Bal}^R$, and $\pi_{\Bal}^I$, 
the interaction part of the Lagrangian between NG modes, $\mathcal{L}_{\mathrm{int}}$, 
which is invariant under the transformations generated by $H_{\bm{s}} \in \Gh$ for any $\bm{s} \in \mathbb{R}^{\bar{r}}$, 
should be a linear combination of the following four terms:
\begin{eqnarray}
&&(\pi_I)^3, \pi_I \Pi_{\Bal}^\ast\Pi_{\Bal} 
 \ \mr{for} \ \Bal \in R_0^{\mr{OD}}
; \nonumber \\
&&\Pi_{\bm{\beta}}^\ast \Pi_{\bm{\gamma}}^\ast \Pi_{\Bal}, 
\Pi_{\bm{\beta}} \Pi_{\bm{\gamma}} \Pi_{\Bal}^\ast
 \ \mr{for} \ \Bal = \bm{\beta} + \bm{\gamma}
.  
\end{eqnarray}
Hence we obtain 
\begin{eqnarray}
\mathcal{L}_{\mathrm{int}} &=& c (\pi_I)^3
 + \pi_I \sum_{\Bal \in R_0^{\mr{OD}}} c_{\Bal} \Pi_{\Bal}^\ast \Pi_{\Bal}
\nonumber \\
&& + \sum_{\Bal, \bm{\beta}, \bm{\gamma} \in R_0^{\mr{OD}}, \Bal = \bm{\beta} + \bm{\gamma}}  \mr{Re}\left(
c_{\Bal, \bm{\beta}, \bm{\gamma}} 
\Pi_{\bm{\beta}}^\ast \Pi_{\bm{\gamma}}^\ast \Pi_{\Bal}
\right)
, \label{eq:intLag1}
\end{eqnarray}
where $c$ and $c_{\Bal}$ are real constants, 
$c_{\Bal, \bm{\beta}, \bm{\gamma}}$ are complex numbers, 
and $\mr{Re}(x)$ denotes the real part of $x$. 
Thus, one phason decays into two phasons $\pi_I$ or 
two oscillatons with the same root vector, 
whereas one oscillaton $\Pi_{\Bal}$ decays into 
one phason $\pi_I$ and one oscillaton $\Pi_{\Bal}$ or 
two oscillatons, $\Pi_{\bm{\beta}}$ and $\Pi_{\bm{\gamma}}$ with $\Bal = \bm{\beta} + \bm{\gamma}$.

Similarly, 
for $\mu$-SB with ODLRO and $\hmu$, 
the interaction Lagrangian between NG modes, $\mathcal{L}_{\mathrm{int}}$, 
can be written up to the third order in $\pi_I$, $\pi_{\Bal}^R$ and $\pi_{\Bal}^I$ as  
\begin{eqnarray}
\mathcal{L}_{\mathrm{int}} &=& c (\pi_I)^3
 + \pi_I \sum_{\Bal \in R_H} c_{\Bal} \Pi_{\Bal}^\ast \Pi_{\Bal}
\nonumber \\
&& + \sum_{\Bal, \bm{\beta}, \bm{\gamma} \in R_H, \Bal = \bm{\beta} + \bm{\gamma}}  \mr{Re}\left(
c_{\Bal, \bm{\beta}, \bm{\gamma}} 
\Pi_{\bm{\beta}}^\ast \Pi_{\bm{\gamma}}^\ast \Pi_{\Bal}
\right)
, \label{eq:intLag2}
\end{eqnarray}
where $c$ and $c_{\Bal}$ are real constants, 
and $c_{\Bal, \bm{\beta}, \bm{\gamma}}$ are complex numbers. 
Thus, one precesson $\Pi_{\Bal}$ decays into 
two phasons $\pi_I$ and a precesson $\Pi_{\Bal}$ or 
two precessons, $\Pi_{\bm{\beta}}$ and $\Pi_{\bm{\gamma}}$ with $\Bal = \bm{\beta} + \bm{\gamma}$.

\section{\label{sec:calculation}Homotopy groups of topological excitations for $\mu$-SB}
Let us now calculate the homotopy groups for $\mu$-SB states to find their topological excitations. 
An element of the first homotopy group $\pi_1(G/H)$ characterizes 
the topological charge of a vortex 
and that of the second homotopy group $\pi_2(G/H)$ characterizes the topological charge of a point defect and that of a skyrmion \cite{Mermin79}. 
Usually, the homotopy groups are calculated separately for individual cases. 
We here show that 
not only the homotopy groups but also the textures of topological excitations can be calculated systematically for $\mu$-SB. 

We first briefly review the theory of an integral lattice and a co-root lattice \cite{Brocker85, Hall15} 
as it is needed for the calculation of the second homotopy group. 
We define integral lattices $L_G$ and $L_H$ for a compact Lie group $G$ and its subgroup $H$ 
and a lattice $L_R(S)$ for a subset $S$ of $R_+$ 
as follows: 
\begin{eqnarray}
L_G &:=& \tbr{\bm{t} \in \mathbb{R}^r | \exp(2 \pi i H_{\bm{t}}) = e, H_{\bm{t}} \in \Gg}
, \\
L_H &:=& \tbr{\bm{t} \in \mathbb{R}^r | \exp(2 \pi i H_{\bm{t}}) = e, H_{\bm{t}} \in \Gh}
, \\
L_R(S) &:=& \mr{Span}_{\mathbb{Z}}\left\{\left.
\frac{2\Bal}{(\Bal, \Bal)} \right| \Bal \in S
\right\}
, \label{eq.co-rootlatiice}
\end{eqnarray}
where $e$ is the identity element of $G$, 
$H_{\bm{t}}$ for $\bm{t} \in \mathbb{R}^r$ is defined by 
\begin{eqnarray}
H_{\bm{t}} = \sum_{b=1}^r t_b H_b
, 
\end{eqnarray}
$\Gg$ and $\Gh$ are the Lie algebras of $G$ and $H$, respectively, 
and $\mr{Span}_{\mathbb{Z}}X$ denotes a vector space spanned by elements of $X$ with integer coefficients:
\begin{eqnarray}
\mr{Span}_{\mathbb{Z}}X := \left\{\left.
\sum_{k=1}^m n_k x_k \right| x_k \in X, n_k \in \mathbb{Z}, m \in \mathbb{N}
\right\}
. 
\nonumber \\
\end{eqnarray}
The vector $2\Bal/(\Bal, \Bal)$ and $L_R(R_+)$ are called the co-root vector (the inverse root vector) of a root vector $\Bal$ 
and the co-root lattice (the inverse root lattice) of $G$, respectively.  

Under the addition of $r$-dimensional vectors, 
$L_G$, $L_H$, and $L_R(S) \ (S \subset R_+)$ form Abelian groups. 
Let us examine this point by discussing an example of $L_G$. 
Let $\bm{t}$ and $\bm{s}$ be elements of $L_G$. 
The sum $\bm{t}+\bm{s}$ satisfies 
\begin{eqnarray}
\exp(2\pi i H_{\bm{t}+\bm{s}}) &=& \exp(2\pi i H_{\bm{t}})\exp(2\pi i H_{\bm{s}})
\nonumber\\
&=& e e = e
, 
\end{eqnarray}
and hence we have $\bm{t}+\bm{s} \in L_G$. 
The identity element of $L_G$ is the zero vector $\bm{0}$ 
and the inverse element of $\bm{t} \in L_G$ is $\bm{-t}$. 
Since $L_R(S)$ is an Abelian group generated by a subset $S$ of $R_+$, 
$L_R(S)$ is an Abelian subgroup of the co-root lattice $L_R(R_+)$. 
Therefore, the coset space $L_R(R_+)/L_R(S)$ becomes an Abelian group. 
Writing the element of the coset space $L_R(R_+)/L_R(S)$ as $[\bm{t}] \ (\bm{t} \in L_R(R_+)$, 
the sum in the coset space $L_R(R_+)/L_R(S)$ is defined as follows:
\begin{eqnarray}
[\bm{t}] + [\bm{s}] :=[\bm{t} + \bm{s}]
. 
\end{eqnarray}
It is known that any co-root vector $2\Bal/(\Bal,\Bal)$ is an element of $L_G$ \cite{Brocker85, Hall15}. 
The co-root vector $2\Bal/(\Bal,\Bal)$ for $\Bal \in R_+ \backslash R_H$ is an element of $L_H$ 
because $H_{\Bal} \in \Gh$ from Theorem \ref{theo:unbroken}, 
where $R_+ \backslash R$ is the set of elements that belong to $R_+$ but not to $R$. 
Therefore, $L_R(R_+ \backslash R_H)$ is a common subgroup of $L_R(R_+)$ and $L_H$, 
and hence the coset space $[L_H \cap L_R(R_+)]/L_R(R_+ \backslash R_H)$ 
forms an Abelian group. 

Then, the following theorem holds. 
\begin{theo}
\label{theo:homotopy}
(1) Provided that $H$ is a connected subgroup of $G$, 
the first homotopy group for $\mu$-SB is given as follows:
\begin{eqnarray}
\pi_1(G/H) &=& \left\{
\begin{array}{ll}
0 
\ \ & \mr{for} \ \mathrm{DLRO}; \\
\mathbb{Z}_l 
\ \ & \mr{for} \ \mathrm{ODLRO \ and} \ \hmu; \\
\mathbb{Z} 
\ \ & \mr{for} \ \mathrm{ODLRO \ and} \ \zmu, \\
\end{array}
\right. 
\label{hom1}
\end{eqnarray}
where $\mathbb{Z}_l = \tbr{g^i | i = 0,1,2, \cdots, l-1}$ is the cyclic group of order $l$ ($g$ is the generator of $\mathbb{Z}_l$), 
$l$ is a positive integer which is uniquely determined from $\hmu$, 
and $\mathbb{Z}_{l=1}$ is a trivial group consisting of the identity element alone. 
Let $\theta \in [0, 2\pi]$ and $O(\theta)$ be the azimuth angle around the vortex 
and the value of the order parameter at the angle $\theta$. 
The vortex with topological charge $g$ represents a vortex 
around which the phase of the order parameter rotates by $2\pi$: 
\begin{eqnarray}
O(\theta) = \mr{e}^{i \theta} O_0
, \label{eq:texvortex}
\end{eqnarray}
where $O_0$ is the value of the order parameter at $\theta=0$. \\
(2) The second homotopy group for $\mu$-SB is 
isomorphic to the coset space constructed from the co-root lattice as follows:
\begin{eqnarray}
\pi_2(G/H) &=& 
\begin{cases}
\frac{L_R(R_+)}{L_R(R_+ \backslash R_H)}
&\ \mr{for} \ \mr{DLRO \ and \ } \hmu; \\
\frac{L_R(R_+)}{L_R(R_+ \backslash R_0^{\mr{D}})}
&\ \mr{for} \ \mr{DLRO \ and \ } \zmu; \\ 
\frac{L_H \cap L_R(R_+)}{L_R(R_+ \backslash R_H)}
&\ \mr{for} \ \mr{ODLRO \ and \ } \hmu; \\
\frac{L_R(R_+)}{L_R(R_+ \backslash R_0^{\mr{OD}})}
&\ \mr{for} \ \mr{ODLRO \ and \ } \zmu.  
\end{cases}
\nonumber\\
\label{hom2}
\end{eqnarray}
Let $\theta\in [0, \pi]$ and $\phi \in [0, 2\pi]$ be the three-dimensional polar coordinates surrounding the point defect, $O(\theta, \phi)$ be the value of the order parameter at $(\theta, \phi)$. 
The texture of the point defect with topological charge $[2\Bal/(\Bal, \Bal)]$ is given by 
\begin{eqnarray}
O(\theta, \phi) &=& 
\exp\left[i \phi \frac{H_{\Bal}}{(\Bal, \Bal)} \right] \circ 
\exp\left[i \theta \frac{E_{\Bal}^I}{(\Bal, \Bal)} \right] \circ O_0
, \nonumber\\
\label{eq:texpointdefect} 
\end{eqnarray}
where $O_0$ and $g \circ O_0 \ (g \in G)$ is the value of the order parameter at $\theta=\phi = 0$ 
and that of the order parameter obtained by the symmetry transformation $g$ from $O_0$, respectively. \\
\end{theo}
The proof of this theorem is given in Appendix \ref{sec:D}. 

Physically, nontrivial elements $\mathbb{Z}_l$ and $\mathbb{Z}$ in Eq.~(\ref{hom1}) 
represent quantum vortices. 
We note that a quantum vortex does not necessarily have an integer charge in a system with internal degrees of freedom \cite{Makela03}. 
Let us examine this point as yet another example of $\mu$-SB. 
Consider the three-dimensional representation of $\bar{\Gg} = \mathfrak{su}(N = 2)$ 
and a phase with ODLRO and $\hmu$. 
This example will appear as a special case of $N=2$ of a $U(N)$-symmetric system in Sec.~\ref{sec:sun}. 
In this representation, 
the order parameter $\langle\bm{\Delta}_s\rangle$ is a $2 \times 2$ symmetric complex matrix 
which transforms under the symmetry transformation of $U(2)$ as follows \cite{Georgi99}: 
\begin{eqnarray}
\langle\bm{\Delta}_s\rangle \mapsto U \langle\bm{\Delta}_s \rangle {}^tU
 \ \mr{for} \ U \in U(2) 
. 
\end{eqnarray}
In $\mu$-SB with ODLRO and $\hmu$, 
the expectation value of $\langle\bm{\Delta}_s\rangle$ is given by 
\begin{eqnarray}
\langle \bm{\Delta}_s \rangle = \langle\bm{\Delta}_0\rangle =
\left( \begin{array}{cc} 1&0\\ 0&0 \end{array} \right)
. 
\end{eqnarray}
Since this representation is the three-dimensional representation of $U(2)$, 
which is equivalent to the three-dimensional representation of $U(1) \times SO(3)$, 
the order parameter $\langle\bm{\Delta}_s\rangle$ is related to 
the three-dimensional complex vector $\langle\bm{\phi}\rangle = (\langle\phi_1\rangle, \langle\phi_0\rangle, \langle\phi_{-1}\rangle)$, 
which is the order parameter of a spin-1 BEC \cite{Ohmi98, Ho98}, as 
\begin{eqnarray}
\langle\bm{\Delta}_s\rangle = \left( \begin{array}{cc} \langle\phi_1\rangle & \langle\phi_0\rangle \\ \langle\phi_0\rangle & \langle\phi_{-1}\rangle \end{array} \right)
. \label{eq.spin1bec}
\end{eqnarray}
The first homotopy group $\pi_1(G/H)$ of this phase is given by 
\begin{eqnarray}
\pi_1(G/H) = \mathbb{Z}_2 = \{ e, g\}
, 
\end{eqnarray}
where $e$ and $g$ are the identity element and the generator of $\mathbb{Z}_2$ with $g^2 = e$. 
From Eq.~(\ref{eq:texvortex}), 
a vortex with topological charge $g$ is given by 
\begin{eqnarray}
\langle\bm{\Delta}_s^{g}\rangle(\theta) = \left(\begin{array}{cc} \mr{e}^{i\theta} & 0 \\  0 & 0 \end{array} \right)
,  
\end{eqnarray}
where $\theta \in [0,2\pi]$ represents the azimuth angle around the vortex. 
We can show $g^2 = e$ as follows. 
Let $\sigma_x, \sigma_y, \sigma_z$ be the Pauli matrices. 
From Eq.~(\ref{eq:texvortex}), 
the texture of the vortex with topological charge $g^2$ is given by 
\begin{eqnarray}
\langle\bm{\Delta}_s^{g^2}\rangle(\theta) &=& \left(\begin{array}{cc} \mr{e}^{2i\theta} & 0 \\  0 & 0 \end{array} \right)
 = U(\theta) \langle\bm{\Delta}_0\rangle {}^tU(\theta)
, \\
U(\theta) &=& \mr{e}^{i \theta \sigma_z}
. 
\end{eqnarray}
By the continuous deformation defined by 
\begin{eqnarray}
\langle\bm{\Delta}_{s,t}\rangle(\theta) &=& U(\theta, t) \langle\bm{\Delta}_0\rangle {}^tU(\theta, t)
, \\
U(\theta, t) &=& \mr{e}^{-i \pi t \frac{\sigma_y}{2}} \mr{e}^{i \theta \frac{\sigma_z}{2}} 
\mr{e}^{i \pi t \frac{\sigma_y}{2}} \mr{e}^{i \theta \frac{\sigma_z}{2}}
 \ \mr{for} \ 0 \le t \le 1
, 
\nonumber\\
\end{eqnarray}
$\langle\bm{\Delta}_{s,t}\rangle$ is transformed from $\langle\bm{\Delta}_{s,t=0}\rangle = \langle\bm{\Delta}_s^{g^2}\rangle$ 
to a uniform order $\langle\bm{\Delta}_{s,t=1}\rangle = \langle\bm{\Delta}_0\rangle$, 
which implies $g^2 = e$. 

From Eqs.~(\ref{eq.co-rootlatiice}) and (\ref{hom2}), 
$\pi_2(G/H)$ is generated by a set of representative elements $\tbr{[2 \Bal/(\Bal, \Bal)] | \Bal \in R_+}$ of the coset space. 
This shows that any point defect in $\mu$-SB can be represented as 
a composite of the point defects with topological charge $[2 \Bal/(\Bal, \Bal)] \ (\Bal \in R_+)$. 
In Eq.~(\ref{hom2}), the coset spaces of $L_R(R_+)$ by their subgroups $L_R(S) \ (S = R_+ \backslash R_H, R_+ \backslash R_0^{\mr{D}}, R_+ \backslash R_0^{\mr{OD}})$ are considered 
instead of the numerators $L_R(R_+)$ or $L_H \cap L_R(R_+)$. 
This is because the element of the denominators $L_R(S)$ of the coset does not give nontrivial topological excitations. 
Let us clarify this point for the case of $\hmu$. 
Since $E_{\Bal}^{R,I}$ and $H_{\Bal}$ for $\Bal \in R_+ \backslash R_H$ are unbroken generators for $\hmu$, 
the successive action of $\exp\left[i \phi \frac{H_{\Bal}}{(\Bal, \Bal)} \right]$ 
and $\exp\left[i \theta \frac{E_{\Bal}^I}{(\Bal, \Bal)} \right]$ leaves the order parameter invariant: 
\begin{eqnarray}
\exp\left[i \phi \frac{H_{\Bal}}{(\Bal, \Bal)} \right] \circ 
\exp\left[i \theta \frac{E_{\Bal}^I}{(\Bal, \Bal)} \right] \circ O_0
 = O_0
\nonumber\\
 \ \mr{for} \ \forall (\theta, \phi) \in [0, \pi] \times [0, 2\pi]
. 
\end{eqnarray}
Therefore, Eq.~(\ref{eq:texpointdefect}) does not give a nontrivial point defect but a uniform order for $\Bal \in R_+ \backslash R_H$ for $\hmu$. 

The point defect in Eq.~(\ref{eq:texpointdefect}) with topological charge $2 \Bal/(\Bal, \Bal)$ 
is similar to the point defect in the FM and the AFM 
in that the former is obtained by replacing $\mathfrak{su}(2)$ spin $\bm{S}$ by the generalized magnetization $\bm{S}_{\Bal}$. 
In fact, let $\bm{M}$ be the magnetization of the FM or that of the sublattice in the AFM. 
A point defect can be described as a hedgehog configuration of $\bm{M}$: 
\begin{eqnarray}
\bm{M}(\theta, \phi) = (\sin\theta\cos\phi, \sin\theta\sin\phi, \cos\theta)
, \label{eq:su2mag}
\end{eqnarray}
where $\theta$ and $\phi$ are 3-dimensional polar coordinates. 
This hedgehog configuration is obtained by 
the successive rotation of the spin around the $y$-axis by angle $\theta$ 
followed by the rotation around the $z$-axis by angle $\phi$: 
\begin{eqnarray}
\bm{M}(\theta, \phi) &=& 
\exp\left(i \phi S_z \right) \circ 
\exp\left(i \theta S_y \right) \circ \bm{M}_0
,
\end{eqnarray}
where $\bm{M}_0 = (0,0,1)$ is the magnetization of the FM or that of the sublattice in the AFM parallel to the $z$-axis. 
Comparing with Eq.~(\ref{eq:texpointdefect}), 
we can see that the point defect in Eq.~(\ref{eq:texpointdefect}) is the generalization 
of the point defect in the FM and the AFM 
obtained by replacing $\bm{S}$ by $\bm{S}_{\Bal}$. 

\section{\label{sec:sun} Application of $\mu$-SB to $U(N)$-symmetric systems} 
In this section, 
we apply $\mu$-SB to $U(N)$-symmetric systems \cite{Fukuhara07, Taie12, Cazalilla09, Gorshkov10}. 
Here, we consider up to the third lowest-dimensional representation of the $\mathfrak{su}(N)$-Lie algebra 
since symmetry broken phases are not necessarily characterized by $\mu$-SB in higher-dimensional representations. 
Up the third lowest-dimensional representation, 
the irreducible representations of the $\mathfrak{su}(N)$-Lie algebra are given by the following three representations, 
the $N$-dimensional representation, the $N(N-1)/2$-dimensional representation, and the $N(N+1)/2$-dimensional representation. 
To make this paper self-contained, 
we briefly review the symmetry transformation, the highest weight, and the set of weight vectors 
of these three representations. 
See Refs.~\cite{Georgi99, Slansky81} for detail on the $\mathfrak{su}(N)$-Lie algebra and its representations.

Let $\tbr{\bm{\nu}_j}_{j=1}^N$ be a set of real $(N-1)$-dimensional vectors 
that satisfy 
\begin{eqnarray}
\sum_{j=1}^N \bm{\nu}_j &=& 0
, \label{eq:sunweightvectors1}\\
(\bm{\nu}_i, \bm{\nu}_j) &=& \delta_{ij} - \frac{1}{N}
. \label{eq:sunweightvectors}
\end{eqnarray}
Defining $\Bal_{i,j}$ by $\Bal_{i,j} := \bm{\nu}_i - \bm{\nu}_j$, 
the set of positive root vectors $R_+$ of the $\mathfrak{su}(N)$-Lie algebra is given by 
\begin{eqnarray}
R_+ &=& \tbr{\Bal_{i,j} | \Bal_{i,j} := \bm{\nu}_i - \bm{\nu}_j, 1 \le i < j \le N}
. \label{eq:sunroots}
\end{eqnarray}
The weight vector $\bm{\nu}_i$ is normalized in Eq.~(\ref{eq:sunweightvectors}) 
so that the magnitude of the root vectors is $\sqrt{2}$. 
For example, $\tbr{\bm{\nu}_j}_{j=1}^N$ for $N=2$ and $3$ are given by 
\begin{eqnarray}
N=2&:& \bm{\nu}_1=\frac{1}{\sqrt{2}}, \bm{\nu}_2=-\frac{1}{\sqrt{2}}
, \\
N=3&:& 
\bm{\nu}_1=\left(\frac{1}{\sqrt{2}}, \frac{1}{\sqrt{6}}\right), 
\bm{\nu}_2=\left(-\frac{1}{\sqrt{2}}, \frac{1}{\sqrt{6}}\right), 
\nonumber\\
&&\bm{\nu}_3=\left(0, \frac{1}{\sqrt{2}}\right)
. 
\end{eqnarray}
For $N=2$ and $N=3$, 
the schematic illustrations are presented in Fig. \ref{fig:weight} (a) and (b), respectively. 

The lowest-dimensional representation is the $N$-dimensional representation. 
The element of this representation is an $N$-dimensional complex vector $\bm{v}$. 
The symmetry transformation of this representation 
acts on $\bm{v}$ as an action of the matrix from the left: 
\begin{eqnarray}
\bm{v} \mapsto U \bm{v}
 \ \mr{for} \ U \in U(N)
. 
\end{eqnarray}
For the ODLRO, $\bm{v}$ is the order parameter of a BEC in degenerate $N$-component bosons. 
The highest weight $\hmu$ and the set of weight vectors $W[\hmu]$ are given by 
\begin{eqnarray}
\hmu &=& \bm{\nu}_1
, \\
W[\bm{\nu}_1] &=& \tbr{\bm{\nu}_j | j=1,2,\cdots, N}
. \label{eq.1weight}
\end{eqnarray}
The weight vector state $| \bm{\nu}_j ) $ is a unit vector whose components vanish except for the $j$-th component: 
\begin{eqnarray}
| \bm{\nu}_j ) = {}^t(0,\ldots,0,{\stackrel{j}{\breve{1}}},0,\ldots,0)
. 
\end{eqnarray}

The second lowest-dimensional representation is the $N(N-1)/2$-dimensional representation. 
The element of this representation is an $N \times N$ complex skew symmetric matrix $\bm{\Delta}_a$. 
In fact, the dimension of the set of $N \times N$ complex skew symmetric matrices is $N(N-1)/2$. 
The symmetry transformation of this representation 
acts on $\bm{\Delta}_a$ as an action of the matrix and that of its transpose from the left and the right, respectively: 
\begin{eqnarray}
\bm{\Delta}_a \mapsto U \bm{\Delta}_a {}^tU
 \ \mr{for} \ U \in U(N)
. \label{eq:transasy}
\end{eqnarray}
For the ODLRO, this order parameter corresponds to that of an $s$-wave superfuild phase 
in degenerate $N$-component fermions in a non-relativistic system \cite{Modawi97, Cherng07, Rapp07, Cazalilla14}. 
In fact, 
let $\tbr{\psi_i}_{i=1}^N$ be the fields of the degenerate $N$-component fermions. 
Since the $N$-components are degenerate 
and the total number of fermion is conserved in a non-relativistic system, 
this system is invariant under the $U(N)$-symmetry transformation: 
\begin{eqnarray}
\psi_i \mapsto U_{ij} \psi_j
 \ \mr{for} \ U \in U(N)
. 
\end{eqnarray}
Here, repeated indices are assumed to be summed over $i=1,\cdots,N$. 
The order parameter of the phase is given by the following $N \times N$ matrix: 
\begin{eqnarray}
\widetilde{\bm{\Delta}}_a = \tbr{\langle \psi_i \psi_j \rangle }_{i,j=1}^N
. 
\end{eqnarray}
The antisymmetric nature of $\widetilde{\bm{\Delta}}_a$ arises from the anticommutation relation of the fermions. 
Under the $U(N)$-symmetric transformation, $\psi_i$ and $\widetilde{\bm{\Delta}}_a$ transform as 
\begin{eqnarray}
\psi_i &\mapsto& U_{ij} \psi_j 
, \\
\left( \widetilde{\bm{\Delta}}_a \right)_{ij} &\mapsto& U_{ik}  \langle \psi_k \psi_l \rangle ({}^tU)_{lj} = \left( U \widetilde{\bm{\Delta}}_a {}^tU\right)_{ij}
\nonumber\\
&& \ \ \ \ \mr{for} \ U \in U(N)
,
\end{eqnarray}
which coincides with the transformation (\ref{eq:transasy}). 
The highest weight $\hmu$ and the set of weight vectors $W[\hmu]$ are given by 
\begin{eqnarray}
\hmu &=& \bm{\nu}_1+\bm{\nu}_2
, \\
W[\bm{\nu}_1+\bm{\nu}_2] &=& \tbr{\bm{\nu}_i+\bm{\nu}_j | 1 \le i < j \le N}
. 
\label{eq.2weight}
\end{eqnarray}
The weight vector state $| \bm{\nu}_i + \bm{\nu}_j ) $ is the skew symmetric matrix 
whose elements vanish except for the $(i,j)$-th and $(j,i)$-th elements: 
\begin{eqnarray}
| \bm{\nu}_i + \bm{\nu}_j ) &=& \bm{\Delta}_a^{(i,j)}
, \\
\left[\bm{\Delta}_a^{(i,j)}\right]_{kl} &=& \delta_{ik}\delta_{jl} - \delta_{il}\delta_{jk}
. 
\end{eqnarray}

The third lowest-dimensional representation is the $N(N+1)/2$-dimensional representation. 
The element of this representation is an $N \times N$ complex symmetric matrix $\bm{\Delta}_s$. 
In fact, the dimension of the set of $N \times N$ complex symmetric matrices is $N(N+1)/2$. 
Similarly to Eq.~(\ref{eq:transasy}), 
the symmetry transformation of this representation 
acts on $\bm{\Delta}_s$ as 
\begin{eqnarray}
\bm{\Delta}_s \mapsto U \bm{\Delta}_s {}^tU
 \ \mr{for} \ U \in U(N)
. \label{eq:transsymm}
\end{eqnarray}
Although the transformation in Eq.~(\ref{eq:transsymm}) coincides with Eq.~(\ref{eq:transasy}), 
$\bm{\Delta}_a$ has the odd parity, ${}^t\bm{\Delta}_a = - \bm{\Delta}_a$, 
and $\bm{\Delta}_s$ the even parity, ${}^t\bm{\Delta}_s = \bm{\Delta}_s$. 
An $N \times N$ skew symmetric (symmetric) matrix $\bm{\Delta}$ transforms into a skew symmetric (symmetric) matrix 
under the transformation $\bm{\Delta} \mapsto U\bm{\Delta}{}^tU \ (U \in U(N))$. 
For the ODLRO, the order parameter $\bm{\Delta}_s$ is related to that of a $p$-wave superfuild phase in degenerate $N$-component fermions \cite{Vollhardt13}. 
In fact, 
let $\tbr{\psi_i}_{i=1}^N$ be the fields of the degenerate $N$-component fermions. 
The order parameter of the $p$-wave superfluid phase is given as follows \cite{Vollhardt13}: 
\begin{eqnarray}
\langle \psi_{i,\bm{k}} \psi_{j,- \bm{k}} \rangle = \sum_{\alpha=x,y,z} k_\alpha \Delta_{\alpha, ij}
.  
\end{eqnarray}
where three $N \times N$ matrices $\bm{\Delta}_\alpha := \tbr{\Delta_{\alpha, ij}}_{i,j=1}^N \ (\alpha = x, y, z)$ 
are symmetric matrices \cite{Vollhardt13}. 
The symmetric nature of $\bm{\Delta}_\alpha$ arises from the anticommutation relation of the fermions 
and the odd parity of the orbital part of the $p$-wave pairing. 
Under the symmetry transformation that mixes the degenerate $N$-components, $\psi_i \mapsto U_{ij} \psi_j$, 
the $N \times N$ symmetric matrix $\bm{\Delta}_s$ transforms according to Eq.~(\ref{eq:transsymm}). 
The highest weight $\hmu$ and the set of weight vectors $W[\hmu]$ are given by 
\begin{eqnarray}
\hmu &=& 2 \bm{\nu}_1
, \\
W[2\bm{\nu}_1] &=& \tbr{\bm{\nu}_i+\bm{\nu}_j | 1 \le i \le j \le N}
. 
\label{eq.3weight}
\end{eqnarray}
The weight vector state $| \bm{\nu}_i + \bm{\nu}_j ) $ is the symmetric matrix 
whose elements vanish except for the $(i,j)$-th and $(j,i)$-th elements: 
\begin{eqnarray}
| \bm{\nu}_i + \bm{\nu}_j ) &=& \bm{\Delta}_s^{(i,j)}
, \\
\left[\bm{\Delta}_s^{(i,j)}\right]_{kl} &=& \delta_{ik}\delta_{jl} + \delta_{il}\delta_{jk}
. 
\end{eqnarray}

\subsection{Classification of $\mu$-SB phases}
We first classify the $\mu$-SB phases for $\bar{\Gg} = \mathfrak{su}(N)$ 
that appear up to the third lowest-dimensional representation. 

\subsubsection{Lowest-dimensional representation}
For $\mu$-SB with ODLRO and $\hmu$, 
the expectation value of the order parameter coincides with the highest weight of this representation in Eq.~(\ref{eq.1weight}): 
\begin{eqnarray}
\langle \bm{\phi} \rangle = | \bm{\nu}_1 ) = (1,0,\cdots, 0)
. 
\end{eqnarray}
The remaining symmetry $H$ of the state is given by 
\begin{eqnarray}
H &:=& \tbr{ U \in U(N) | U | \bm{\nu}_1 ) =  | \bm{\nu}_1 )}
\nonumber\\ 
&=& \tbr{ U \in U(N) | U_{1j} = U_{j1} = \delta_{j1} \ (j=1,2,\cdots,N)}
\nonumber\\ 
&\simeq& U(N-1)
, 
\end{eqnarray}
where $\simeq$ represents the group isomorphism. 
Thus, $H$ is a connected group. 
In this representation, the Casimir invariant $C_2^{\bar{\Gg}} (\vev)$ coincides with $|\vev|^4$ 
and hence $C_2^{\bar{\Gg}} (\vev) \ne 0$ so long as the order parameter has a nonzero expectation value. 
Therefore the pair of ODLRO and $\bm{\mu}_0$ is absent in this representation. 

We next consider the case of DLRO. 
There are $N$-fold degenerate states in each site which are referred to as color or flavor. 
For $\mu$-SB with DLRO and $\hmu$, 
the mean-field ground state is given by 
\begin{eqnarray}
| GS \rangle = \bigotimes_{i \in L} | \bm{\nu}_1 )_i 
. 
\end{eqnarray} 
This is the mean-field of an $SU(N)$-ferromagnet \cite{Cazalilla09}. 
The unitary operator $\widehat{U}$ associated with the symmetry transformation $U\in U(N)$ 
leaves $| GS \rangle$ unchanged up to a global phase factor 
if and only if $U$ leaves the weight vector $| \bm{\nu}_1 )$ on each sites unchanged up to a global phase: 
\begin{eqnarray}
U | \bm{\nu}_1 ) = \mr{e}^{i\phi} | \bm{\nu}_1 )
 \ \mr{for} \ \exists \phi \in \mathbb{R}
. 
\end{eqnarray}
This condition is equivalent to the condition that 
the first row and the first column of $U$ vanish except for a diagonal element: 
\begin{eqnarray}
U_{1j} = U_{j1} = 0 \ \mr{for} \ j=2,3,\cdots,N
. 
\end{eqnarray}
The remaining symmetry $H$ of the state is given by 
\begin{eqnarray}
H &:=& \tbr{ U \in U(N) | U_{1j} = U_{j1} = 0 \ (j=2,3,\cdots,N)}
\nonumber\\ 
&\simeq& U(1) \times U(N-1)
, \label{eq:Hcase1}
\end{eqnarray}
and hence $H$ is a connected group. 
For the case of $\zmu$, 
from Eq.~(\ref{eq.casimirunit}) the expectation value of the Casimir invariant within the unit cell $u$ is given by 
\begin{eqnarray}
C_2^{\bar{\Gg}} = \left\lVert \sum_{i \in u} \bm{\mu}_i \right\rVert^2
. 
\label{eq.muz}
\end{eqnarray}
From Eq.~(\ref{eq:sunweightvectors1}), 
the right-hand side of Eq.~(\ref{eq.muz}) vanishes 
when the unit cell consists of $N$-different weight vectors, $\{ \bm{\mu}_i \}_{i \in u} = \{ \bm{\nu}_j \}_{j=1}^N$. 
Thus, the ground state is given by 
\begin{eqnarray}
| GS \rangle = \bigotimes_{u \in \mathcal{U}} \bigotimes_{i \in u} | \bm{\mu}_i )_i 
, \\
\{ \bm{\mu}_i \}_{i \in u} = \{ \bm{\nu}_j \}_{j=1}^N
. \label{eq.DLROz1}
\end{eqnarray} 
This is the mean-field ground state of the so-called $N$-color density wave ($N$-CDW) phase \cite{Tsunetsugu06, Lauchli06, Cai13b, Nataf14}. 
This state is a generalization of the $SU(2)$-antiferromagnet to a general $SU(N)$-spin system. 
In the former, two states, a spin-up state and a spin-down state, constitute a unit cell, 
while in the latter $N$ states do. 
The unitary operator $\widehat{U}$ associated with the symmetry transformation $U\in U(N)$ 
leaves $| GS \rangle$ unchanged up to a global phase factor 
if and only if $U$ leaves all of the $N$-different weight vectors $| \bm{\nu}_i )$ within each unit cell unchanged up to a global phase. 
In other words, the condition 
\begin{eqnarray}
U | \bm{\nu}_i ) = \mr{e}^{i\phi} | \bm{\nu}_i )
 \ \mr{for} \ \exists \phi \in \mathbb{R}
, 
\end{eqnarray}
must be satisfied for all of $N$-weight vectors $\{\bm{\nu}_j\}_{j=1}^N$. 
This condition is satisfied if and only if $U$ is a diagonal matrix. 
Thus, the remaining symmetry $H$ of the state is given by 
\begin{eqnarray}
H &:=& \tbr{ U \in U(N) | U =\mr{diag}(\mr{e}^{i\phi_1}, \mr{e}^{i\phi_2}, \cdots, \mr{e}^{i\phi_N}), \ \phi_i \in \mathbb{R}}
\nonumber\\ 
&\simeq& U(1)^N 
, \label{eq:H}
\end{eqnarray}
and hence $H$ is a connected group.

\subsubsection{Second lowest-dimensional representation}
For $\mu$-SB with ODLRO and $\hmu$, 
the expectation value of the order parameter is given from Eqs.~(\ref{eq.musb1}) and (\ref{eq.weight2}) by 
\begin{eqnarray}
\bm{\Delta}_a = \left| \bm{\nu}_1 + \bm{\nu}_2 \right) = \bm{\Delta}_a^{(1,2)}
. 
\end{eqnarray}
The remaining symmetry $H$ of the state is determined by a straightforward calculation as 
\begin{eqnarray}
H &:=& \tbr{ U \in U(N) | U \bm{\Delta}_a^{(1,2)} {}^tU = \bm{\Delta}_a^{(1,2)}}
\nonumber\\ 
&=& \{ U \in U(N) | U_{11}U_{22} - U_{12}U_{21} = 1, 
\nonumber\\
&&\ \ \ U_{ij} = U_{ji} = 0 \ (i=1,2, j=3,4,\cdots,N)\}
\nonumber\\ 
&\simeq& SU(2) \times U(N-2)
,  
\end{eqnarray}
and hence $H$ is a connected group. 
On the other hand, $\mu$-SB with ODLRO and $\zmu$ does not necessarily exist for arbitrary $N$. 
The same point is discussed in the derivation of the mean fields of $\mu$-SB in Sec.~\ref{sec:definition}. 
The set $W[\bm{\nu}_1 + \bm{\nu}_2]$ does not necessarily contain the zero-weight vector \cite{Georgi99}. 
In fact, $\bm{\nu}_i + \bm{\nu}_j = \bm{0} \ (1 \le i < j \le N)$ 
only when the following conditions are met: $N=2$, $\bm{\nu}_i = \bm{\nu}_1$, and $\bm{\nu}_j =\bm{\nu}_2$. 
For $N$ greater than $2$, 
$W[\bm{\nu}_1 + \bm{\nu}_2]$ does not contain the zero-weight vector. 

We next consider the case of DLRO. 
There are $N(N-1)/2$-degenerate states labeled by a weight vector $\bm{\mu} \in W[\bm{\nu}_1+ \bm{\nu}_2]$ in each site. 
For $\mu$-SB with DLRO and $\hmu$, 
the mean-field ground state is given from Eqs.~(\ref{eq.musb3}) and (\ref{eq.weight2}) by 
\begin{eqnarray}
| GS \rangle = \bigotimes_{i \in L} | \bm{\nu}_1 + \bm{\nu}_2 )_i 
. 
\end{eqnarray} 
Similarly to the DLRO and $\hmu$ in the lowest-dimensional representation, 
the unitary operator $\widehat{U}$ associated with the symmetry transformation $U\in U(N)$ 
leaves $| GS \rangle$ unchanged up to a global phase factor 
if and only if $U$ leaves the weight vector $| \bm{\nu}_1 + \bm{\nu}_2 )$ on each site unchanged up to a global phase: 
\begin{eqnarray}
&&U | \bm{\nu}_1 + \bm{\nu}_2 ) = \mr{e}^{i\phi} | \bm{\nu}_1 + \bm{\nu}_2 )
 \ \mr{for} \ \exists \phi \in \mathbb{R}
\nonumber\\
&\Leftrightarrow& U_{ij} = U_{ji} = 0 \ (i=1,2, j=3,4,\cdots,N)
. 
\end{eqnarray}
Thus, the remaining symmetry $H$ of the state is given by 
\begin{eqnarray}
H &:=& \{ U \in U(N) | 
\nonumber\\
&&\ \ \ U_{ij} = U_{ji} = 0 \ (i=1,2, j=3,4,\cdots,N)\}
\nonumber\\ 
&\simeq& U(2) \times U(N-2)
,  
\end{eqnarray}
and hence $H$ is a connected group. 
For the case of $\zmu$, 
the right-hand side of Eq.~(\ref{eq.muz}) vanishes 
when the set of weight vectors $\{ \bm{\mu}_i \}_{i \in u}$on the unit cell $u$ is 
\begin{eqnarray}
\begin{cases}
\{ \bm{\mu}_i \}_{i \in u} = \{ \bm{\nu}_{2j-1} + \bm{\nu}_{2j} \}_{j=1}^{N/2}
& \ \mr{for \ even} \ N
; \\
\{ \bm{\mu}_i \}_{i \in u} = \{ \bm{\nu}_{j} + \bm{\nu}_{j+1} \}_{j=1}^{N-1} \cup \{ \bm{\nu}_1 + \bm{\nu}_N \}
& \ \mr{for \ odd} \ N
.  
\end{cases}
\nonumber\\
\end{eqnarray}
For even $N$, $N/2$-sites within each unit cell are sufficient because 
we have from Eq.~(\ref{eq:sunweightvectors1}) 
\begin{eqnarray}
\sum_{j=1}^{N/2} (\bm{\nu}_{2j-1} + \bm{\nu}_{2j}) = \sum_{j=1}^N \bm{\nu}_j = 0
. 
\end{eqnarray}
On the other hand, $(N-1)/2$-sites within the unit cell are not sufficient 
because the sum of the weight vectors within each set $\{ \bm{\nu}_{2j-1} + \bm{\nu}_{2j} \}_{j=1}^{(N-1)/2}$ is 
\begin{eqnarray}
\sum_{j=1}^{(N-1)/2} (\bm{\nu}_{2j-1} + \bm{\nu}_{2j}) = \sum_{j=1}^{N-1} \bm{\nu}_j = -\bm{\nu}_N \ne 0
. 
\end{eqnarray}
We have to consider the unit cell with $N$-sites. 
In fact, the set $\{ \bm{\nu}_{j} + \bm{\nu}_{j+1} \}_{j=1}^{N-1} \cup \{ \bm{\nu}_1 + \bm{\nu}_N \}$ satisfies 
\begin{eqnarray}
\sum_{j=1}^{N-1}(\bm{\nu}_{j} + \bm{\nu}_{j+1} ) + ( \bm{\nu}_1 + \bm{\nu}_N ) = 2 \sum_{j=1}^N \bm{\nu}_j = 0
\end{eqnarray}
Thus, the ground state is given by 
\begin{eqnarray}
&&| GS \rangle = \bigotimes_{u \in \mathcal{U}} \bigotimes_{i \in u} | \bm{\mu}_i )_i 
, \\
&&\begin{cases}
\{ \bm{\mu}_i \}_{i \in u} = \{ \bm{\nu}_{2j-1} + \bm{\nu}_{2j} \}_{j=1}^{N/2}
& \ \mr{for \ even} \ N
; \\
\{ \bm{\mu}_i \}_{i \in u} = \{ \bm{\nu}_{j} + \bm{\nu}_{j+1} \}_{j=1}^{N-1} \cup \{ \bm{\nu}_1 + \bm{\nu}_N \}
& \ \mr{for \ odd} \ N
. 
\end{cases}
\nonumber\\
\label{eq.DLROz2}
\end{eqnarray} 
The remaining symmetry $H$ of the state can be calculated in a manner similar to the derivation of Eq.~(\ref{eq:H}). 
A unitary matrix $U$ is included in $H$ 
if and only if $U| \bm{\mu}_i ) = \mr{e}^{i\phi} | \bm{\mu}_i ) \ (\exists \phi \in \mathbb{R})$ for all of the weight vectors within each unit cell. 
For even $N$, $U| \bm{\nu}_{2j-1} + \bm{\nu}_{2j} ) = \mr{e}^{i\phi} | \bm{\nu}_{2j-1} + \bm{\nu}_{2j} ) \ (\exists \phi \in \mathbb{R})$ 
is satisfied when both the $(2j-1)$ and the $(2j)$-th columns and the $(2j-1)$ and the $(2j)$-th rows vanish except for the $(2j-1,2j-1), (2j-1,2j), (2j,2j-1)$, and $(2j,2j)$-elements. 
Therefore, 
$U$ is block-diagonalized into a direct product of $2 \times 2$ matrices 
and we obtain 
\begin{eqnarray}
H &=& 
\left\{U \in U(N) \left| U = \bigoplus_{j=1}^{N/2} U_i \ (U_i \in U(2)) \right. \right\}
, \\
&\simeq& \left[ U(2) \right]^{N/2}
. 
\end{eqnarray}
For odd $N$ and the set of weight vectors $\{ \bm{\nu}_{j} + \bm{\nu}_{j+1} \}_{j=1}^{N-1} \cup \{ \bm{\nu}_1 + \bm{\nu}_N \}$, 
a unitary matrix $U$ is included in $H$ 
if and only if $U$ is a diagonal matrix. 
Therefore, 
we obtain 
\begin{eqnarray}
H &:=& \{ U \in U(N) | U =\mr{diag}(\mr{e}^{i\phi_1}, \mr{e}^{i\phi_2}, \cdots, \mr{e}^{i\phi_N}) \ \phi_i \in \mathbb{R}\}
\nonumber\\ 
&\simeq& U(1)^N 
, 
\end{eqnarray}
For both even and odd $N$, 
$H$ is a connected group. 

\subsubsection{Third lowest-dimensional representation}
For $\mu$-SB with ODLRO and $\hmu$, 
the expectation value of the order parameter is given by 
\begin{eqnarray}
\bm{\Delta}_s = | 2\bm{\nu}_1) = \bm{\Delta}_s^{(1,1)}
. 
\end{eqnarray}
The remaining symmetry $H$ of the state is given by 
\begin{eqnarray}
H &:=& \tbr{ U \in U(N) | U \bm{\Delta}_s^{(1,1)} {}^tU = \bm{\Delta}_s^{(1,1)}}
\nonumber\\ 
&=& \tbr{ U \in U(N) | U_{1j} = U_{j1} = 0 \ (j=1,2,\cdots,N)}
\nonumber\\ 
&\simeq& U(N-1)
,  
\end{eqnarray}
and hence $H$ is a connected group. 
Similar to the case of the second lowest-dimensional representation, 
$\mu$-SB with ODLRO and $\zmu$ does not necessarily exist for arbitrary $N$. 
The set $W[2\bm{\nu}_1]$ contains the weight vector $\bm{\nu}_i + \bm{\nu}_j = \bm{0} \ (1 \le i \le j \le N)$ 
only when the following conditions are met: $N=2$, $\bm{\nu}_i = \bm{\nu}_1$ and $\bm{\nu}_j =\bm{\nu}_2$ \cite{Georgi99}. 
For $N$ greater than $2$, 
$W[2 \bm{\nu}_1]$ does not contain the zero-weight vector. 
For the ODLRO and $\zmu$ with $N=2$, $H$ is not a connected group. 
In fact, the order parameter of this phase and $H$ are given by 
\begin{eqnarray}
\bm{\Delta}_s &=& \sigma_x
, \\
H &=& \tbr{\mr{e}^{i t} I_2 | t \in \mathbb{R}} \rtimes \left\{I_2, \mr{e}^{i\frac{\pi}{2}} \sigma_y \right\} 
\nonumber\\
&=& U(1) \rtimes \mathbb{Z}_2
, 
\end{eqnarray}
where $N \rtimes N^\prime$ is a semidirect product whose product is given by 
\begin{eqnarray}
(n, h) \rtimes (n^\prime, h^\prime) = (n hn^\prime h^{-1}, h h^\prime)
\nonumber\\
 \ \mr{for} \ 
\forall n, n^\prime \in N, \forall h, h^\prime \in N^\prime
. 
\end{eqnarray}

We next consider the case of DLRO. 
There are $N(N+1)/2$ states labeled by a weight vector $\bm{\mu} \in W[2\bm{\nu}_1]$ in each site. 
For $\mu$-SB with DLRO and $\hmu$, 
the mean-field ground state is given by 
\begin{eqnarray}
| GS \rangle = \bigotimes_{i \in L} | 2\bm{\nu}_1)_i 
. 
\end{eqnarray} 
The remaining symmetry $H$ of the state is calculated in a manner similar to Eq.~(\ref{eq:Hcase1}) as 
\begin{eqnarray}
H &:=& \tbr{ U \in U(N) | U \bm{\Delta}_s^{(1,1)} {}^tU = \mr{e}^{i\phi} \bm{\Delta}_s^{(1,1)} \ \mr{for} \ \exists \phi \in \mathbb{R}}
\nonumber\\ 
&=& \tbr{ U \in U(N) | U_{1j} = U_{j1} = 0 \ (j=2,3,\cdots,N)}
\nonumber\\ 
&\simeq& U(1) \times U(N-1)
, 
\end{eqnarray}
and hence $H$ is a connected group. 
For the case of $\zmu$, 
the ground state and the remaining symmetry $H$ of the state 
is obtained in a manner similar to the case of the DLRO and $\zmu$ 
for the lowest-dimensional representation. 
the right-hand side of Eq.~(\ref{eq.muz}) vanishes when 
the set of weight vectors $\{ \bm{\mu}_i \}_{i \in u}$ within the unit cell $u$ are
\begin{eqnarray}
\{ \bm{\mu}_i \}_{i \in u} = \{ 2 \bm{\nu}_j \}_{j=1}^N
. 
\end{eqnarray}
Thus, the ground state is given by 
\begin{eqnarray}
&&| GS \rangle = \bigotimes_{u \in \mathcal{U}} \bigotimes_{i \in u} | \bm{\mu}_i )_i 
, \\
&& \{ \bm{\mu}_i \}_{i \in u} = \{ 2 \bm{\nu}_j \}_{j=1}^N
. \label{eq.DLROz3}
\end{eqnarray} 
The remaining symmetry $H$ of the state can be calculated in a manner similar to the derivation of Eq.~(\ref{eq:H}). 
A unitary matrix $U$ is included in $H$ if and only if $U$ is a diagonal matrix. 
Thus, the remaining symmetry $H$ of the state is given by 
\begin{eqnarray}
H &:=& \tbr{ U \in U(N) | U =\mr{diag}(\mr{e}^{i\phi_1}, \mr{e}^{i\phi_2}, \cdots, \mr{e}^{i\phi_N}), \ \phi_i \in \mathbb{R}}
\nonumber\\ 
&\simeq& U(1)^N 
, 
\end{eqnarray}
and hence $H$ is a connected group.

\subsection{Numbers of NG modes}
We next calculate the number of NG modes in $\mu$-SB phases classified in the previous subsection. 
From the quadratic part of effective Lagrangians in Eq.~(\ref{effLag}), 
the numbers of type-1 and type-2 NG modes, $n_1$ and $n_2$, are given by
\begin{eqnarray}
(n_1, n_2)&=& 
\left\{
\begin{array}{ll}
(0, |R_H|)
\ \ & \mr{for} \ \mathrm{DLRO \ and} \ \hmu; \\
(2|R_0^{\mr{D}}|, 0)
\ \ & \mr{for} \ \mathrm{DLRO \ and} \ \zmu; \\
(1, |R_H|)
\ \ & \mr{for} \ \mathrm{ODLRO \ and} \ \hmu; \\
(2|R_0^{\mr{OD}}| + 1, 0)
\ \ & \mr{for} \ \mathrm{ODLRO \ and} \ \zmu, 
\end{array}
\right. 
\label{eq.counting}
\nonumber\\
\end{eqnarray}
where $|X|$ denotes the number of elements in the set $X$.

\subsubsection{Lowest-dimensional representation}
Combining the set $W[\hmu]$ of the weight vectors in Eq.~(\ref{eq.1weight}) and 
the set $\{ \bm{\mu}_i \}_{i \in u}$ of the weight vectors in the unit cell $u$ in Eq.~(\ref{eq.DLROz1}), 
the sets $R_H$ and $R_0^{\mr{D}}$ are given from Eqs.~(\ref{eq.RH}) and (\ref{eq.R0D}) by 
\begin{eqnarray}
R_H &=& \tbr{\Bal_{1,j} | 2 \le j \le N}
, \\
R_0^{\mr{D}} &=& \tbr{\Bal_{i,j} | 1 \le i < j \le N}  =R_+
. 
\label{eq.1root}
\end{eqnarray}
Substituting these equations into Eq.~(\ref{eq.counting}), 
we obtain 
\begin{eqnarray}
(n_1, n_2) &=& 
\left\{
\begin{array}{ll}
(0, N-1) 
\ \ & \mr{for} \ \mathrm{DLRO \ and} \ \hmu; \\
(N(N-1),0)
\ \ & \mr{for} \ \mathrm{DLRO \ and} \ \zmu; \\
(1, N-1) 
\ \ & \mr{for} \ \mathrm{ODLRO \ and} \ \hmu. 
\end{array}
\right. 
\nonumber\\
\end{eqnarray}

\subsubsection{Second lowest-dimensional representation}
Combining the set $W[\hmu]$ of the weight vectors in Eq.~(\ref{eq.2weight}) and 
the set $\{ \bm{\mu}_i \}_{i \in u}$ of the weight vectors in the unit cell $u$ in Eq.~(\ref{eq.DLROz2}), 
the sets $R_H$ and $R_0^{\mr{D}}$ are given from Eqs.~(\ref{eq.RH}) and (\ref{eq.R0D}) by 
\begin{eqnarray}
R_H &=& \tbr{\Bal_{i,j} | i=1,2, \ 3 \le j \le N}
, \\
R_0^{\mr{D}} &=& R_+ \backslash \tbr{\Bal_{2j-1,2j} | j =1,2,\cdots, N/2} \ \mr{for} \ \mr{even} \ N
, \nonumber\\
\\
R_0^{\mr{D}} &=& \tbr{\Bal_{i,j} | 1 \le i < j \le N}  = R_+ \ \mr{for} \ \mr{odd} \ N
. 
\label{eq.2root}
\end{eqnarray}
For $N=2$, there exists the $\mu$-SB phase with ODLRO and $\zmu$. 
From Eq.~(\ref{eq.R0OD}), the set $R_0^{\mr{OD}}$ in this phase is empty:
\begin{eqnarray}
R_0^{\mr{OD}} = \emptyset
, 
\end{eqnarray}
where $\emptyset$ denotes the empty set. 
Substituting the above three equations into Eq.~(\ref{eq.counting}), 
we obtain 
\begin{eqnarray}
&&(n_1, n_2) \nonumber\\
&=& 
\left\{
\begin{array}{ll}
(0, 2(N-2)) 
\ \ & \mr{for} \ \mathrm{DLRO \ and} \ \hmu; \\
(N(N-2), 0) 
\ \ & \mr{for} \ \mathrm{DLRO,} \ \zmu, \mr{and} \ \mr{even} \ N; \\
(N(N-1), 0) 
\ \ & \mr{for} \ \mathrm{DLRO,} \ \zmu, \mr{and} \ \mr{odd} \ N; \\
(1, 2(N-2)) 
\ \ & \mr{for} \ \mathrm{ODLRO \ and} \ \hmu; \\
(1, 0) 
\ \ & \mr{for} \ \mathrm{ODLRO,} \ \zmu, \mr{and} \ N=2. 
\end{array}
\right. 
\nonumber\\
\end{eqnarray}

\subsubsection{Third lowest-dimensional representation}
Combining the set $W[\hmu]$ of the weight vectors in Eq.~(\ref{eq.3weight}) and 
the set $\{ \bm{\mu}_i \}_{i \in u}$ of the weight vectors in the unit cell $u$ in Eq.~(\ref{eq.DLROz3}), 
the sets $R_H$ and $R_0^{\mr{D}}$ are given from Eqs.~(\ref{eq.RH}) and (\ref{eq.R0D}) by 
\begin{eqnarray}
R_H &=& \tbr{\Bal_{1,j} |  2 \le j \le N}
, \\
R_0^{\mr{D}} &=& \tbr{\Bal_{i,j}| 1 \le i < j \le N}  = R_+
. 
\label{eq.3root}
\end{eqnarray}
For $N=2$, there exists a $\mu$-SB phase with ODLRO and $\zmu$. 
The set $R_0^{\mr{OD}}$ in this phase is given from the definition of $R_0^{\mr{OD}}$ in Eq.~(\ref{eq.R0OD}) as 
\begin{eqnarray}
R_0^{\mr{OD}} = \tbr{\Bal_{1,2}}
. 
\end{eqnarray}
Substituting the above equations into Eq.~(\ref{eq.counting}), 
we obtain 
\begin{eqnarray}
&&(n_1, n_2) \nonumber\\
&=& 
\left\{
\begin{array}{ll}
(0, N-1) 
\ \ & \mr{for} \ \mathrm{DLRO \ and} \ \hmu; \\
(N(N-1),0)
\ \ & \mr{for} \ \mathrm{DLRO \ and} \ \zmu; \\
(1, N-1) 
\ \ & \mr{for} \ \mathrm{ODLRO \ and} \ \hmu; \\
(3, 0) 
\ \ & \mr{for} \ \mathrm{ODLRO,} \ \zmu, \mr{and} \ N=2. \\
\end{array}
\right. 
\nonumber\\
\end{eqnarray}

\subsection{Homotopy groups of topological excitations}
Finally, we calculate the first and second homotopy groups for $\mu$-SB phases. 
Since $H$ is a connected group in all the cases except for the ODLRO and $\zmu$ with $N=2$ in the third lowest-dimensional representation, 
we can apply Theorem \ref{theo:homotopy} except for this case. 

\subsubsection{Lowest-dimensional representation}
For the DLRO, from Eq.~(\ref{hom1}), we obtain 
\begin{eqnarray}
\pi_1(G/H) = 0
. 
\end{eqnarray}
For the DLRO and $\hmu$, 
we obtain from $R_+$ in Eq.~(\ref{eq:sunroots}) 
\begin{eqnarray}
L_R(R_+) &=& 
\mr{Span}_{\mathbb{Z}}\left\{\left. \frac{2\Bal}{(\Bal, \Bal)} \right| \Bal \in R_+ \right\}
\nonumber\\
&=&\mr{Span}_{\mathbb{Z}}\left\{ \Bal_{i,j} | 1 \le i \le N \right\}
. \label{eq.deriving}\\
L_R(R_+ \backslash R_H) &=& 
\mr{Span}_{\mathbb{Z}}\left\{ \Bal_{i,j} | 2 \le i \le N \right\}
, 
\end{eqnarray}
In deriving the first equality of Eq.~(\ref{eq.deriving}), 
we use Eq.~(\ref{eq:sunroots}) and $(\Bal_{i,j}, \Bal_{i,j}) = 2$ for any $i,j$. 
Therefore we obtain from Theorem \ref{theo:homotopy}
\begin{eqnarray}
\pi_2(G/H) &=& \frac{L_R(R_+)}{L_R(R_+ \backslash R_H)}
\nonumber\\
&=& \mr{Span}_{\mathbb{Z}} \tbr{\Bal_{1,2}} \simeq \mathbb{Z}
. 
\label{eq.2ndhom1Dh}
\end{eqnarray}
For the DLRO and $\zmu$, since $R_0^{\mr{D}} =R_+$, we obtain from Theorem \ref{theo:homotopy}
\begin{eqnarray}
\pi_2(G/H) &=& L_R(R_+) 
\nonumber\\
&=& \left\{ \left. \sum_{j, k=1}^N m_{jk} \Bal_{j,k} \right| m_{jk} \in \mathbb{Z} \right\}
\nonumber\\
&=& \left\{ \left. \sum_{j=1}^N m_j \bm{\nu}_j \right| m_j \in \mathbb{Z}, \sum_{j=1}^N m_j = 0 \right\}
\nonumber\\
&\simeq& \mathbb{Z}^{N-1}
. 
\label{eq.2ndhom1Dz}
\end{eqnarray}

For the ODLRO and $\hmu$, 
it is easier to calculate $\pi_1(G/H)$ and $\pi_2(G/H)$ directly 
rather than using Theorem \ref{theo:homotopy} 
because the order parameter manifold is isomorphic to a higher-dimensional sphere: 
\begin{eqnarray}
G/H= U(N)/U(N-1) = S^{2N-1}
. 
\end{eqnarray}
Since $N \ge 2$, we obtain the following two equations from the standard results of homotopy groups \cite{Mermin79}: 
\begin{eqnarray}
\pi_1(G/H) &=& \pi_1(S^{2N-1}) = 0
, \\
\pi_2(G/H) &=& \pi_2(S^{2N-1}) = 0
. 
\end{eqnarray}

\subsubsection{Second lowest-dimensional representation}
For the DLRO, from Eq.~(\ref{hom1}), we obtain 
\begin{eqnarray}
\pi_1(G/H) = 0
. 
\end{eqnarray}
For the DLRO and $\hmu$, from Theorem \ref{theo:homotopy} and Eq.~(\ref{eq.2root}), we obtain 
\begin{eqnarray}
\pi_2(G/H) &=& \frac{L_R(R_+)}{L_R(R_+ \backslash R_H)}
\nonumber\\
&=& \mr{Span}_{\mathbb{Z}} \tbr{\Bal_{2,3}} \simeq \mathbb{Z}
. 
\end{eqnarray}
For the DLRO, $\zmu$ and even $N$, 
from Theorem \ref{theo:homotopy} and Eq.~(\ref{eq.DLROz2}), 
we obtain 
\begin{eqnarray}
L_R(R_+ \backslash R_0^{\mr{OD}}) &=& 
\left\{\left. \sum_{j=1}^{N/2} m_j \Bal_{2j-1,2j} \right| m_j \in \mathbb{Z} \right\}
, \nonumber\\
\\
\pi_2(G/H) &=& \frac{L_R(R_+)}{L_R(R_+ \backslash R_0^{\mr{OD}})}
\nonumber\\
&=& \left\{ \left. \sum_{j=1}^{N/2-1} m_j \Bal_{2j,2j+1} \right|m_j \in \mathbb{Z} \right\}
\nonumber\\
&\simeq& \mathbb{Z}^{\frac{N}{2}-1}
. 
\end{eqnarray}
For the DLRO, $\zmu$ and odd $N$, we obtain the same result $\pi_2(G/H) = \mathbb{Z}^{N-1}$ 
as in the case of the lowest-dimensional representation since $R_0^{\mr{D}}$ is the same in both cases 
of the lowest-dimensional and second lowest-dimensional representations.

For the ODLRO and $\hmu$, 
we can prove $\pi_1(G/H) = 0$ as follows. 
To show this, it is sufficient to show that the following vortex-like texture analogous to Eq.~(\ref{eq:texvortex}) 
can be deformed into a uniform order: 
\begin{eqnarray}
\langle \bm{\Delta}_a \rangle = \mr{e}^{i\theta} \bm{\Delta}_a^{(1,2)}
. \label{eq.trivialvortex}
\end{eqnarray}
Here, $\theta \in [0,2\pi]$ is the azimuth angle around the vortex-like object. 
Let $\lambda_z$ and $\lambda_y$ be two matrices defined by 
\begin{eqnarray}
\lambda_z &=& \left( \begin{array}{ccc} 1&0&0 \\ 0&0&0 \\ 0&0&-1 \end{array} \right)
, \\
\lambda_y &=& \left( \begin{array}{ccc} 0&0&-i \\ 0&0&0 \\ i&0&0 \end{array} \right)
, 
\end{eqnarray}
and define $N \times N$ matrices $\widetilde{\lambda}_x$ and $\widetilde{\lambda}_y$ by 
a direct product of $\lambda_z$ and $\lambda_y$ with the identity matrix $I_{N-3}$ of size $(N-3)$, respectively: 
\begin{eqnarray}
\widetilde{\lambda}_z &=& \lambda_z \oplus I_{N-3}
, \\
\widetilde{\lambda}_y &=& \lambda_y \oplus I_{N-3}
.  
\end{eqnarray}
By using $\widetilde{\lambda}_z$, Eq.~(\ref{eq.trivialvortex}) can be written as 
\begin{eqnarray}
\langle \bm{\Delta}_a \rangle &=& U(\theta) \bm{\Delta}_a^{(1,2)} {}^tU(\theta)
, \\
U(\theta) &=& \mr{e}^{i \theta \widetilde{\lambda}_z}
. 
\end{eqnarray}
Consider the continuous deformation defined by 
\begin{eqnarray}
\langle\bm{\Delta}_{a,t}\rangle(\theta) &=& U(\theta, t) \langle\bm{\Delta}_a^{(1,2)}\rangle {}^tU(\theta, t)
, \\
U(\theta, t) &=& \mr{e}^{-i \pi t \frac{\widetilde{\lambda}_y}{2}} \mr{e}^{i \theta \frac{\widetilde{\lambda}_z}{2}} 
\mr{e}^{i \pi t \frac{\widetilde{\lambda}_y}{2}} \mr{e}^{i \theta \frac{\widetilde{\lambda}_z}{2}}
\nonumber\\
&& \ \mr{for} \ 0 \le t \le 1
. 
\end{eqnarray}
The unitary matrix $U(\theta, t)$ satisfies
\begin{eqnarray}
U(\theta, t=0) &=& \exp(i \theta \widetilde{\lambda}_z) = U(\theta)
, \\
U(\theta, t=1) &=&
\mr{e}^{-i \pi \frac{\widetilde{\lambda}_y}{2}} \mr{e}^{i \theta \frac{\widetilde{\lambda}_z}{2}} 
\mr{e}^{i \pi \frac{\widetilde{\lambda}_y}{2}} \mr{e}^{i \theta \frac{\widetilde{\lambda}_z}{2}}
\nonumber\\
&=&
\mr{e}^{-i \theta \frac{\widetilde{\lambda}_z}{2}} \mr{e}^{i \theta \frac{\widetilde{\lambda}_z}{2}} = I_N
, 
\end{eqnarray}
where $I_N$ denotes the identity matrix of size $N$. 
In the third line, 
we use the relation 
\begin{eqnarray}
\mr{e}^{-i \pi \frac{\widetilde{\lambda}_y}{2}} \widetilde{\lambda}_z \mr{e}^{i \pi \frac{\widetilde{\lambda}_y}{2}} = 
- \widetilde{\lambda}_z
. 
\end{eqnarray}
Therefore, $\langle\bm{\Delta}_{a,t}\rangle$ is transformed from Eq.~(\ref{eq.trivialvortex}) 
to a uniform order $\langle\bm{\Delta}_{a,t=1}\rangle = \bm{\Delta}_a^{(1,2)}$, 
which implies the triviality of the vortex-like texture in Eq.~(\ref{eq.trivialvortex}). 
From Eq.~(\ref{eq.2root}), 
we obtain 
\begin{eqnarray}
&&L_H 
\nonumber\\
&=& \left\{ \left. t \Bal_{1,2} + \sum_{j,k=3}^N m_{jk} \Bal_{j,k} \right|
 t, m_{jk} \in \mathbb{R}, 3 \le i < j \le N \right\}
, \nonumber\\
\\
&&L_H \cap L_R(R_+) 
\nonumber\\
&=& 
\mr{Span}_{\mathbb{Z}} \left[ \{\Bal_{i,j} | 3 \le i < j \le N \} 
\cup \tbr{\Bal_{1,2}} \right]
\nonumber\\
&=& L_R(R_+ \backslash R_H)
. 
\end{eqnarray}
Therefore, we obtain $\pi_2(G/H) = 0$ from Eq.~(\ref{hom2}). 
For the ODLRO and $\zmu$ with $N=2$, 
by using Eq.~(\ref{hom1}) and substituting $R_0^{\mr{OD}} = \emptyset$ into Eq.~(\ref{hom2}), 
we obtain 
\begin{eqnarray}
\pi_1(G/H) &=& \mathbb{Z}
, \\
\pi_2(G/H) &=& 0
. 
\end{eqnarray}

\subsubsection{Third lowest-dimensional representation}
For the DLRO, from Eq.~(\ref{hom1}) we obtain 
\begin{eqnarray}
\pi_1(G/H) = 0
. 
\end{eqnarray}
For the DLRO and $\hmu$, we obtain the same result $\pi_2(G/H) = \mathbb{Z}$ 
as in the case of the lowest-dimensional representation since $R_H$ coincides in both cases 
of the lowest-dimensional and third lowest-dimensional representations. 
For the DLRO and $\zmu$, we obtain the same result $\pi_2(G/H) = \mathbb{Z}^{N-1}$ 
as in the case of the lowest-dimensional representation since $R_0^{\mr{D}}$ coincides in both cases 
of the lowest-dimensional and third lowest-dimensional representations.

For the ODLRO and $\hmu$, 
we can prove $\pi_1(G/H) = \mathbb{Z}_2$ in a manner similar to the discussion in Sec.~\ref{sec:calculation}. 
Let $g$ be the generator of $\mathbb{Z}_2$. 
The vortex with topological charge $g^2$ is described as 
\begin{eqnarray}
\bm{\Delta}_s(\theta) = \mr{e}^{2 i\theta}\bm{\Delta}_s^{(1,1)}
,  
\end{eqnarray}
where $\theta \in [0,2\pi]$ is the azimuth angle around the vortex. 
We can prove $g^2 = e$ in a manner similar to the case of $N=2$ in Sec.~\ref{sec:calculation} 
by replacing Pauli matrices $\sigma_i (i = x,y,z)$ into 
\begin{eqnarray}
\widetilde{\sigma}_i := \sigma_i \oplus I_{N-2} 
(i = x,y,z) 
. 
\end{eqnarray}
For the ODLRO and $\hmu$, we obtain the same result $\pi_2(G/H) = 0$ 
as in the case of the lowest-dimensional representation since $L_H$ and $R_H$ coincide in both cases 
of the lowest-dimensional and third lowest-dimensional representations. 
For the ODLRO and $\zmu$ with $N=2$, 
Eq.~(\ref{hom1}) is not applicable because $H$ is not a connected group. 
From the correspondence with the spin-1 BEC in Eq.~(\ref{eq.spin1bec}), 
this phase coincides with the polar phase in the spin-1 BEC \cite{Ohmi98, Ho98}. 
The first and second homotopy groups of this phase are given as follows \cite{Leonhardt00, Stoof01}: 
\begin{eqnarray}
\pi_1(G/H) &=& \mathbb{Z}
, \\
\pi_2(G/H) &=& \mathbb{Z}
. 
\end{eqnarray}

We list the results obtained in this section in Table \ref{tab:table1} 
together with the examples of the classified phases. 
Examples with $\mathfrak{g} = \mathfrak{u}(1) \oplus \mathfrak{so}(3)$ are included 
because $\mathfrak{so}(3)$ is isomorphic to $\mathfrak{su}(2)$. 
From the fifth column of Table \ref{tab:table1}, 
we can see that a large class of symmetry broken phases are described in terms of $\mu$-SB. 

\begin{table*}
\caption{\label{tab:table1}
Classification of $\mu$-symmetry breaking in systems without Lorentz invariance. 
The Lie algebras of the systems are assumed to take the form of $\mathfrak{g} = \mathfrak{u}(1) \oplus \mathfrak{su}(N)$ with $N \ge 2$. 
Here, $\{ \bm{\nu}_i \}_{i=1}^N$ is the set of weight vectors in an $N$-dimensional representation of $\mathfrak{su}(N)$, 
which satisfies Eqs.~(\ref{eq:sunweightvectors1}) and (\ref{eq:sunweightvectors}). 
The first column shows the irreducible representation of the order parameter $\vev$ for ODLRO 
and that of the field of particles on each site of the lattice $L$ for DLRO. 
The second column shows whether the system is characterized 
by $\hmu$ or $\zmu$ and by ODLRO or DLRO, respectively. 
The expectation value of the order parameter for ODLRO can be written in the form of Eq.~(\ref{eq.musb1}) or Eq.~(\ref{eq.musb2}), 
while the ground state for DLRO can be written in the form of Eq.~(\ref{eq.musb3}) or Eq.~(\ref{eq.musb4}). 
The row with $\zmu$ and ODLRO appears only for $N=2$. 
For the representation of the $\hmu = \bm{\nu}_1$ case, the pair of $\zmu$ and ODLRO is absent 
since this representation is the lowest-dimensional one. 
The third column $(n_1, n_2)$ shows the numbers of type-1 and type-2 NG modes. 
The fourth column $(\pi_1, \pi_2)$ lists the first and second homotopy groups of $G/H$. 
FM, AFM, CDW, SF, and VBS stand for ferromagnet, antiferromagnet, color density wave, superfluid, and valence bond solid, respectively. 
In the row with $2\bm{\nu}_1$, $\zmu$, and DLRO, 
the upper (lower) row corresponds to the case of even (odd) $N$. 
}
\begin{ruledtabular}
\begin{tabular}{ll||ll|l}
\textrm{representation $\hmu$} & \textrm{classification} & $(n_1, n_2)$ & $(\pi_1, \pi_2)$ & \textrm{example} \\
\colrule
\hline
$\bm{\nu}_1$ & ODLRO, $\hmu$ & $(1,N-1)$ & $(0,0)$ & $SU(N)$-FM BEC \\
\cline{2-5}
            & DLRO, \ \ $\hmu$  & $(0,N-1)$& $(0,\mbZ)$ & spin-1/2 FM, $SU(N)$-FM \cite{Cazalilla09}\\
\cline{2-5}
            & DLRO, \ \ $\zmu$   & $(N(N-1),0)$& $(0,\mbZ^{N-1})$ & spin-1/2 AFM, $N$-CDW \cite{Tsunetsugu06, Lauchli06, Cai13b, Nataf14}\\
\hline
$\bm{\nu}_1 + \bm{\nu}_2$ & ODLRO, $\hmu$ & $(1,2(N-2))$& $(0,0)$ & $s$-wave SF in 3-component fermion
\cite{Modawi97, Cherng07, Rapp07}\\
\cline{2-5}
                        & ODLRO, $\zmu$ & $(1,0)$& $(\mbZ,0)$ & $s$-wave SF in 2-component fermion
\cite{Bardeen57}\\
\cline{2-5}
                        & DLRO, \ \ $\hmu$ & $(0,2(N-2))$& $(0,\mbZ)$ &  \\
\cline{2-5}
                        & DLRO, \ \ $\zmu$ & $(N(N-2),0)$& $(0,\mbZ^{\frac{N}{2}-1})$ & 
VBS in $SU(4)$-spin model \cite{Assaad05, Corboz11}\\
\cline{3-5}
              &  & $(N(N-1),0)$& $(0,\mbZ^{N-1})$ & \\
\hline
$2\bm{\nu}_1$ & ODLRO, $\hmu$& $(1,N-1)$& $(\mbZ_2,0)$ & FM phase in spin-1 BEC \cite{Ohmi98, Ho98}\\
\cline{2-5}
              & ODLRO, $\zmu$ & $(3,0)$& $(\mbZ, \mbZ)$ & polar phase in spin-1 BEC \cite{Ohmi98, Ho98}\\
\cline{2-5}
              & DLRO, \ \ $\hmu$   & $(0,N-1)$& $(0,\mbZ)$ & spin-1 FM \\
\cline{2-5}
              & DLRO, \ \ $\zmu$   & $(N(N-1),0)$& $(0,\mbZ^{N-1})$ & spin-1 AFM \\
\end{tabular}
\end{ruledtabular}
\end{table*}

\section{\label{sec:application} Discussion on the case of higher-dimensional representation}
So far we have confined our discussions to low-dimensional representations. 
We next turn to the case of a higher-dimensional representation. 
In the case of ODLRO in a higher-dimensional representation, 
there appears more than one Casimir invariant in the energy functional. 
Due to the competition between these Casimir invariants, 
the phases that arise from the minimization of the energy functional 
are, in general, described by neither $\mu$-SB nor inert states, 
where an inert state is a state in which the order parameter is independent of the coupling constants \cite{Michel71, Michel80}. 
In this section, we focus on the case of a higher-dimensional representation 
in which two Casimir invariants appear in the energy functional. 
In this case, many of the ground states are described by inert states 
despite the competition between Casimir invariants. 
Let us examine this point by discussing examples of spin-2 BECs \cite{Koashi00, Ciobanu00, Ueda02} 
and spin-1 color superconductors \cite{Schmitt05, Brauner08}. 

\subsection{Spin-2 BEC}
First, we consider the example of spin-2 BECs. 
As we will see below, all of the ground states are inert states. 
The symmetry group of the system is $U(1) \times SO(3)$. 
The order parameter of spin-2 BEC is a five-dimensional complex vector 
\begin{eqnarray}
\vev = {}^t( \langle \phi_2 \rangle, \langle \phi_1 \rangle, \langle \phi_0 \rangle, \langle \phi_{-1} \rangle, \langle \phi_{-2} \rangle)
, 
\end{eqnarray} 
and the Cartan generator of $SO(3)$ is the $S_z$-operator defined by 
\begin{eqnarray}
S_z = 
\left(
\begin{array}{ccccc}
2&0&0&0&0\\
0&1&0&0&0\\
0&0&0&0&0\\
0&0&0&-1&0\\
0&0&0&0&-2
\end{array}
\right)
. 
\end{eqnarray}
The three eigenstates of the $S_z$-operator 
\begin{eqnarray}
\vev_H &=& {}^t(1,0,0,0,0)
, \\
\vev_Z &=& {}^t(0,0,1,0,0)
, \\ 
\vev_L &=& {}^t(0,0,0,0,1)
, 
\end{eqnarray}
are the highest-weight, the zero-weight, and the lowest-weight states, respectively. 
For $\mathfrak{g}= \mathfrak{u}(1) \oplus \mathfrak{so}(3)$, the lowest-weight state is obtained by applying $\pi$-rotation around the $x$-axis to the highest-weight state. 
The mean-field energy functional can be constructed from 
the norm $|\vev|$ of the order parameter and 
the Casimir invariants of $SO(5)$ and $SO(3)$ as 
\begin{eqnarray}
V(\vev) &=& - c |\vev|^2 + c^\prime_0 |\vev|^4 + c^\prime_1 C_2^{\mathfrak{so}(5)}(\vev) 
\nonumber \\
&&+ c^\prime_2 C_2^{\mathfrak{so}(3)}(\vev)
, 
\end{eqnarray}
where $C_2^{\mathfrak{so}(5)}(\vev)$ and $C_2^{\mathfrak{so}(3)}(\vev)$ 
are the Casimir invariants of $SO(5)$ and $SO(3)$, respectively \cite{Uchino08}. 
Here $C_2^{\mathfrak{so}(5)}(\vev)$ is related to the spin-singlet pair amplitude 
\begin{eqnarray}
A_{00}(\bm{\phi}) := \frac{1}{\sqrt{5}}
[\phi_2 \phi_{-2} - \phi_1 \phi_{-1} + (\phi_0)^2]
\end{eqnarray}
as 
\begin{eqnarray}
C_2^{\mathfrak{so}(5)}(\vev) = |\vev|^4-5|A_{00}(\vev)|^2
. 
\end{eqnarray}
In the energy functional, there are two competing Casimir invariants, namely $C_2^{\mathfrak{so}(5)}(\vev)$ and $C_2^{\mathfrak{so}(3)}(\vev)$. 
By minimizing the energy functional, 
the following four phases are obtained \cite{Koashi00, Ciobanu00, Ueda02}:
\begin{eqnarray}
\begin{cases}
\mr{ferromagnetic \ phase:} \ 
\vev = \vev_H = {}^t(1,0,0,0,0)
\\
\ \ \ \mr{for} \ c^\prime_1 < 0 \ \mr{and} \ c^\prime_2 < 0
; \\
\mr{cyclic \ phase:} \ 
\vev = {}^t\left(\frac{1}{2}, 0, \frac{i}{\sqrt{2}}, 0, \frac{1}{2} \right)
\\
\ \ \ \mr{for} \ c^\prime_1 < 0 \ \mr{and} \ c^\prime_2 > 0
; \\
\mr{uniaxial \ nematic \ phase:} \ 
\vev = \vev_Z = {}^t(0,0,1,0,0)
\\
\ \ \ \mr{for} \ c^\prime_1 > 0
; \\
\mr{biaxial \ nematic \ phase:} \ 
\vev = {}^t\left(\frac{1}{\sqrt{2}}, 0, 0, 0, \frac{1}{\sqrt{2}} \right)
\\
\ \ \ \mr{for} \ c^\prime_1 > 0
, 
\end{cases}
\nonumber\\
\end{eqnarray}
where the order parameters are normalized such that $|\vev| = 1$. 
We note that the uniaxial nematic and biaxial nematic phases are energetically degenerate at the mean-field level. 

While the ferromagnetic phase and the uniaxial nematic phase are described by $\mu$-SB with $\hmu$ and $\zmu$, respectively, 
the cyclic phase and the biaxial nematic phase are not. 
However, both the cyclic phase and the biaxial nematic phase are inert states. 
Moreover, they are both described by linear combinations of the highest-weight, zero-weight, and lowest-weight states 
with simple ratios between the coefficients, $1$ and $\sqrt{2} i$. 

\subsection{Spin-1 color superconductor}
Next, we consider the example of spin-1 color superconductors. 
As we will see below, the four ground states are obtained from the minimization of the energy functional 
and three of them are inert while one of them is not. 
Color superconducting phases are the superconducting phases 
in which Cooper pairs formed by quarks are condensed \cite{Alford08}. 
Since each quark field $q^c_{f,s}$ has three internal degrees of freedom, flavor $f$, spin $s$, and color $c$, 
the resulting Cooper pair has these three internal degrees of freedom. 
In the spin-1 color superconducting phases, 
quarks form a Cooper pair in a single flavor, a spin $SO(3)$-triplet, and a color $SU(3)$-antitriplet channel \cite{Schmitt05, Brauner08}. 
The single flavor and the spin $SO(3)$-triplet imply that 
the Cooper pair does not have an internal degree of freedom in the flavor but has the spin 1, respectively. 
The color $SU(3)$-antitriplet implies that 
the Cooper pair has three color charge, anti-red, anti-blue, and anti-green. 
Let $\Delta_{c,l}$ be the field of the Cooper pair 
with color $c \ (=1,2,3)$ and the spin direction parallel to the $l \ ( = x,y,z)$-axis, respectively, 
where colors $1,2$, and $3$ denote the color charge anti-red, anti-blue, and anti-green, respectively. 
The term ``anti" implies that 
$\Delta_{c,l}$ transforms in the conjugate representation of the three-dimensional representation of $SU(3)$: 
\begin{eqnarray}
\Delta_{c,l} &\mapsto& (U_{c c^\prime})^\ast \Delta_{c,l} 
\nonumber\\
 \ \mr{when} \ 
q^c_{f,s} &\mapsto& U_{c c^\prime} q^{c^\prime}_{f,s} \ \ (U \in SU(3)
. \label{eq.symmetyrCSC1}
\end{eqnarray}
Under the spin rotation, 
$\bm{\Delta}_{c,l}$ transforms in the vector representation of $SO(3)$: 
\begin{eqnarray}
\bm{\Delta}_{c,l} \mapsto  R_{l l^\prime} \bm{\Delta}_{c,l^\prime}
 \ \mr{for} \ R \in SO(3)
. \label{eq.symmetyrCSC2}
\end{eqnarray}
Also, there is the $U(1)$-symmetry associated with the baryon-number conservation 
which acts on the order parameter $\bm{\Delta}$ as 
\begin{eqnarray}
\bm{\Delta} \mapsto \mr{e}^{i\phi} \bm{\Delta} 
\ \mr{for} \ \phi \in \mathbb{R}
. \label{eq.symmetyrCSC3}
\end{eqnarray}
Based on the above discussions, 
the order parameter of the spin-1 color superconducting phase is given by a $3 \times 3$ complex matrix 
\begin{eqnarray}
\bm{\Delta} = \tbr{ \Delta_{c,l} | c = 1,2,3, l = x,y,z}
, 
\end{eqnarray}
and the symmetry group $G$ is $G = U(3) \times SO(3)$ 
which consists of three symmetries, 
the color $SU(3)$-symmetry, the spin $SO(3)$-symmetry, and the $U(1)$-symmetry associated with the baryon number conservation, respectively. 
Combining Eqs.~(\ref{eq.symmetyrCSC1}), (\ref{eq.symmetyrCSC2}), and (\ref{eq.symmetyrCSC3}), 
the order parameter $\bm{\Delta}$ transforms under $G$ as 
\begin{eqnarray}
\bm{\Delta} \mapsto U^\ast \bm{\Delta} {}^tR
\ \mr{for} \ U \in U(3), R \in SO(3)
. \label{eq.actionCSC}
\end{eqnarray}
We note that the system has a combined symmetry of $U(3)$ and $SO(3)$, 
resulting in the two Casimir invariants in the energy functional. 
The Cartan generators of the Lie algebra of $G$ consist of three generators; 
two generators, $\lambda_3$ and $\lambda_8$, of $\mathfrak{su}(3)$ 
and one generator, $S_z$, of $\mathfrak{so}(3)$. 
They are defined as 
\begin{eqnarray}
\lambda_3 &=& 
\left( \begin{array}{ccc}
1&0&0\\0&-1&0\\0&0&0
\end{array} \right)
, \ \lambda_8 = 
{1 \over \sqrt{3}}
\left( \begin{array}{ccc}
1&0&0\\0&1&0\\0&0&-2
\end{array} \right)
, \\
S_z &=& 
\left( \begin{array}{ccc}
0&i&0\\-i&0&0\\0&0&0
\end{array} \right)
. 
\end{eqnarray}
We note that from Eq.~(\ref{eq.actionCSC}) 
the actions of $\lambda_3, \lambda_8$, and $S_z$ on the order parameter commute 
because the $SU(3)$-group and its generators act on the order parameter $\bm{\Delta}$ from the left, 
while the $SO(3)$-group and its generators act on it from the right. 
The mean-field energy functional of spin-1 color superconductors 
can be constructed from 
the Hilbert-Schmit norm $\sqrt{\mr{Tr}(\bm{\Delta}\bm{\Delta}^\dagger)}$ of the matrix $\bm{\Delta}$ and 
the Casimir invariants of $SU(3)$ and $SO(3)$ as 
\begin{eqnarray}
V(\bm{\Delta}) &=& - \bar{c} \mr{Tr}(\bm{\Delta}\bm{\Delta}^\dagger) + 
\bar{c}_0 \left[\mr{Tr}(\bm{\Delta}\bm{\Delta}^\dagger)\right]^2 + 
\bar{c}_1 C_2^{\mathfrak{su}(3)}(\bm{\Delta}) 
\nonumber \\
&&+ \bar{c}_2 C_2^{\mathfrak{so}(3)}(\bm{\Delta})
. 
\end{eqnarray}
Here, $C_2^{\mathfrak{su}(3)}(\bm{\Delta})$ and $C_2^{\mathfrak{so}(3)}(\bm{\Delta})$ 
are the Casimir invariants of $SU(3)$ and $SO(3)$ defined by 
\begin{eqnarray}
C_2^{\mathfrak{su}(3)}(\bm{\Delta}) &=& 
\sum_{a=1}^8 \left[\mr{Tr}(\bm{\Delta} \lambda_a \bm{\Delta}^\dagger)\right]^2
, \\
C_2^{\mathfrak{so}(3)}(\bm{\Delta}) &=& 
\sum_{a=1}^3 \left[\mr{Tr}(\bm{\Delta} S_a \bm{\Delta}^\dagger)\right]^2
, 
\end{eqnarray}
where $\{ \lambda_a \}_{a=1}^8$ and $\{ S_a \}_{a=1}^3$ are 
the set of the Gell-Mann matrices of $\mathfrak{su}(3)$ \cite{Georgi99} and 
the set of the generators of the vector representation of $\mathfrak{so}(3)$ 
defined by $(S_a)_{bc} = i \epsilon^{abc}$, 
respectively. 
In the energy functional, there are two competing Casimir invariants, namely $C_2^{\mathfrak{su}(3)}(\bm{\Delta})$ and $C_2^{\mathfrak{so}(3)}(\bm{\Delta})$. 
These Casimir invariants are related to the quartic invariants 
$\mr{Tr}(\bm{\Delta} \bm{\Delta}^\dagger\bm{\Delta} \bm{\Delta}^\dagger)$ 
and $\mr{Tr}\left[\bm{\Delta} {}^t\bm{\Delta} (\bm{\Delta} {}^t\bm{\Delta})^\dagger\right]$ 
used in Ref.~\cite{Brauner08} 
as 
\begin{eqnarray}
C_2^{\mathfrak{su}(3)}(\bm{\Delta}) &=& 
2 \mr{Tr}(\bm{\Delta} \bm{\Delta}^\dagger\bm{\Delta} \bm{\Delta}^\dagger)
 -{2 \over 3} \left[\mr{Tr}(\bm{\Delta} \bm{\Delta}^\dagger)\right]^2
, \nonumber\\ \\
C_2^{\mathfrak{so}(3)}(\bm{\Delta}) &=& 
\mr{Tr}(\bm{\Delta} \bm{\Delta}^\dagger\bm{\Delta} \bm{\Delta}^\dagger)
 - \mr{Tr}\left[\bm{\Delta} {}^t\bm{\Delta} (\bm{\Delta} {}^t\bm{\Delta})^\dagger \right]
. \nonumber\\  
\end{eqnarray}
From the analysis of the quartic invariants in Ref.~\cite{Brauner08}, 
they satisfy the following inequalities 
\begin{eqnarray}
0 &\le& C_2^{\mathfrak{su}(3)}(\bm{\Delta}) \le {4\over 3}, 
\\ 
0 &\le& C_2^{\mathfrak{so}(3)}(\bm{\Delta}) \le 1
\end{eqnarray}
for a $3 \times 3$ matrix normalized as $\mr{Tr}(\bm{\Delta}\bm{\Delta}^\dagger) = 1$. 
By minimizing the energy functional, 
the following four phases are obtained \cite{Brauner08}:
\begin{eqnarray}
&&\mr{A \ phase:} \ 
\bm{\Delta}_A = 
{1\over 2} 
\left( \begin{array}{ccc}
1&i&0\\-i&1&0\\0&0&0
\end{array} \right)
\nonumber\\
&& \ \ \mr{for} \ 2\bar{c}_1+\bar{c}_2< 0 \ \mr{and} \ \bar{c}_2 < 0
, \\
&&\mr{polar \ phase:} \ 
\bm{\Delta}_P = 
\left( \begin{array}{ccc}
0&0&0\\0&0&0\\0&0&1
\end{array} \right)
\nonumber\\
&& \ \ \mr{for} \ \bar{c}_1 < 0 \ \mr{and} \ \bar{c}_2 > 0
, \\
&&\mr{color-spin-locked \ phase:} \ 
\bm{\Delta}_C = 
{1 \over \sqrt{3}}
\left( \begin{array}{ccc}
1&0&0\\0&1&0\\0&0&1
\end{array} \right)
\nonumber\\
&& \ \ \mr{for} \ \bar{c}_1 > 0 \ \mr{and} \ \bar{c}_1 + \bar{c}_2 > 0
, \label{eq.spin1CSC}\\
&&\epsilon\mr{ \ phase:} \ 
\bm{\Delta}_\epsilon = 
\left( \begin{array}{ccc}
\epsilon_1&i\epsilon_1&0\\-i\epsilon_1&\epsilon_1&0\\0&0&\epsilon_2
\end{array} \right)
\nonumber\\
&& \ \ \mr{for} \ 2\bar{c}_1+\bar{c}_2 > 0 \ \mr{and} \ \bar{c}_1 + \bar{c}_2 < 0
, 
\end{eqnarray}
where $\epsilon_1$ and $\epsilon_2$ are the constant values defined as 
\begin{eqnarray}
\epsilon_1 = {1 \over 2} \sqrt{{2\bar{c}_1 \over 4\bar{c}_1 + \bar{c}_2}}
, \ \ \epsilon_2 = \sqrt{{2\bar{c}_1 + \bar{c}_2 \over 4\bar{c}_1 + \bar{c}_2}}
. 
\end{eqnarray}
The order parameters are normalized such that $\mr{Tr}(\bm{\Delta}\bm{\Delta}^\dagger) = 1$. 

Let us analyze these phases from the viewpoint of $\mu$-SB, the inert state, and the Casimir invariants. 
Since the symmetry group of this system is no longer a simple Lie group, 
we generalize the concept of $\mu$-SB for ODLRO to the case 
in which the symmetry group of the system is a semisimple Lie group. 
Since a semisimple Lie algebra can be decomposed into the direct sum of 
the one-dimensional commutative Lie algebras and simple Lie algebras, 
we can assign the Casimir invariants to each simple Lie algebra. 
When the order parameter is a simultaneous eigenstate of all of the Cartan generators and 
each Casimir invariant is minimized or maximized for the state, 
we refer to such a symmetry breaking as $\mu$-symmetry breaking. 
Among the four ground states, 
the A phase, the polar phase, and the color-spin-locked phase are inert states, 
while the $\epsilon$ phase is not. 
The order parameter of the A phase is a simultaneous eigenstate of $\lambda_3, \lambda_8$, and $S_z$: 
\begin{eqnarray}
\lambda_3 \bm{\Delta}_A = \bm{\Delta}_A, \ 
\lambda_8 \bm{\Delta}_A = {\bm{\Delta}_A \over \sqrt{3}}, \ 
\bm{\Delta}_A S_z = \bm{\Delta}_A
. 
\end{eqnarray}
We note from Eq.~(\ref{eq.actionCSC}) that the generator $S_z$ of $\mathfrak{so}(3)$ acts on the order parameter from the right. 
In the A phase, both $C_2^{\mathfrak{su}(3)}(\bm{\Delta})$ and $C_2^{\mathfrak{so}(3)}(\bm{\Delta})$ 
are maximized: 
\begin{eqnarray}
C_2^{\mathfrak{su}(3)}(\bm{\Delta}_A) = {4\over 3}, \ 
C_2^{\mathfrak{so}(3)}(\bm{\Delta}_A) = 1
.
\end{eqnarray}
Therefore, the A phase is described by $\mu$-SB. 
The order parameter of the polar phase is a simultaneous eigenstate of $\lambda_3, \lambda_8$, and $S_z$, 
\begin{eqnarray}
\lambda_3 \bm{\Delta}_P = {2 \over \sqrt{3}}\bm{\Delta}_P, \ 
\lambda_8 \bm{\Delta}_P = 0, \ 
\bm{\Delta}_P S_z = 0
, 
\end{eqnarray}
and $C_2^{\mathfrak{su}(3)}(\bm{\Delta})$ is maximized 
, whereas $C_2^{\mathfrak{so}(3)}(\bm{\Delta})$ is minimized: 
\begin{eqnarray}
C_2^{\mathfrak{su}(3)}(\bm{\Delta}_P) = {4\over 3}, \ 
C_2^{\mathfrak{so}(3)}(\bm{\Delta}_P) = 0
.
\end{eqnarray}
Therefore, the polar phase is described by $\mu$-SB. 
The color-spin-locked phase is an inert state but is not $\mu$-SB. 
However, we can see from Eq.~(\ref{eq.spin1CSC}) 
that the ratios between the components are all simple numbers 
similarly to the case of the spin-2 BECs; they are all one. 
In the color-spin-locked phase, 
both $C_2^{\mathfrak{su}(3)}(\bm{\Delta})$ and $C_2^{\mathfrak{so}(3)}(\bm{\Delta})$ 
are minimized: 
\begin{eqnarray}
C_2^{\mathfrak{su}(3)}(\bm{\Delta}_C) = 0, \ 
C_2^{\mathfrak{so}(3)}(\bm{\Delta}_C) = 0
.
\end{eqnarray}
The $\epsilon$ phase is not an inert state. 
This phase is an intermediate phase between the A phase and the polar phase. 
In the limit $\bar{c}_1/\bar{c}_2 \to -1/2$ \ ($\bar{c}_1/\bar{c}_2 \to 0$), 
it coincides with the A phase \ (the polar phase). 
In the $\epsilon$ phase, 
the Casimir invariants takes intermediate values between their minimum and maximum: 
\begin{eqnarray}
C_2^{\mathfrak{su}(3)}(\bm{\Delta}_\epsilon) &=& {4 \over 3} - { 8\bar{c}_1(2\bar{c}_1 +\bar{c}_2) \over (4\bar{c}_1 +\bar{c}_2)^2}, \\ 
C_2^{\mathfrak{so}(3)}(\bm{\Delta}_\epsilon) &=& \left({ 2 \bar{c}_1 \over 4\bar{c}_1 +\bar{c}_2} \right)^2
.
\end{eqnarray}
In spin-1 color superconductors, 
a non-inert state emerges as a consequence of the competition between different Casimir invariants.

\section{\label{sec:conclusion}Conclusion}
In conclusion, 
we have proposed a Lie-algebraic approach to systematically finding mean fields of quantum many-body systems 
on the basis of the dynamical symmetry. 
The mean fields of $\mu$-symmetry breaking is derived through the minimization of the energy functional constructed from the Casimir invariants. 
We have introduced a concept of $\mu$-symmetry breaking 
as a phase that is characterized by a weight vector in the representation of the Lie algebra. 
For $\mu$-SB, 
the quadratic part of an effective Lagrangian of NG modes is block-diagonalized as in Eq.~(\ref{effLag}) 
in terms of the Cartan canonical form. 
In $\mu$-SB there appear three types of NG modes as listed in Table \ref{tab:table3}. 
Also, homotopy groups of topological excitations are calculated systematically for $\mu$-SB as summarized in Eqs.~(\ref{hom1}) and (\ref{hom2}). 
The textures of NG modes and topological excitations are described in terms of the generalized magnetization $\bm{S}_{\Bal}$. 
By applying $\mu$-SB to a $U(N)$-symmetric system, 
we have demonstrated that $\mu$-SB involves a large class of symmetry broken phases as listed in Table \ref{tab:table1}. 

In Sec.~\ref{sec:application}, 
we have seen from the examples of spin-2 BECs and spin-1 color superconductors that 
many of the ground states obtained by the minimization of the energy functional are inert, 
despite the fact that there is a competition between different Casimir invariants. 
Moreover, these states are described by linear combinations of weight vectors with simple ratios between the coefficients. 
The physics behind this fact is yet to be fully understood and merits further study.

\begin{acknowledgments}
This work was supported by
KAKENHI Grant No. 26287088 from the Japan Society for the Promotion of Science,
a Grant-in-Aid for Scientific Research on Innovation Areas ``Topological Quantum Phenomena'' (KAKENHI Grant No. 22103005),
the Photon Frontier Network Program from MEXT of Japan,
and the Mitsubishi Foundation.
S. H. acknowledges support from JSPS (Grant
No. 16J03619) and through the Advanced Leading Graduate
Course for Photon Science (ALPS). 
\end{acknowledgments}

\appendix*

\section{\label{sec:D}Proof of Theorem \ref{theo:homotopy} on the homotopy groups in $\mu$-SB}
In this Appendix, 
we prove Theorem \ref{theo:homotopy} for $\mu$-SB 
on the basis of the theory of an integral lattice and a co-root lattice \cite{Brocker85, Hall15}.

Let $r$ and $R_+$ be the rank and the set of positive roots of a compact Lie group $G$. 
We define an integral lattice $L_G$ for a compact Lie group $G$ and 
a lattice $L_R(S)$ for a subset $S$ of $R_+$ as follows \cite{Brocker85, Hall15}: 
\begin{eqnarray}
L_G &:=& \tbr{\bm{t} \in \mathbb{R}^r | \exp(2 \pi i H_{\bm{t}}) = e}
, \\
L_R(S) &:=& \mr{Span}_{\mathbb{Z}}\left\{\left.
\frac{2\Bal}{(\Bal, \Bal)} \right| \Bal \in S
\right\}
,
\end{eqnarray}
where $\mr{Span}_{\mathbb{Z}}X$ denotes a vector space spanned by elements of $X$ with integer coefficients:
\begin{eqnarray}
\mr{Span}_{\mathbb{Z}}X := \left\{\left.
\sum_{k=1}^{n^\prime} n_k x_k \right| x_k \in X, n_k \in \mathbb{Z}, n^\prime \in \mathbb{N}
\right\}
. 
\nonumber \\
\end{eqnarray}
Under a addition of $r$-dimensional vectors, 
$L_G$ and $L_R(S)$ of $R_+$ become Abelian groups for any subset $S$. 
It is known that $L_R(S)$ is an Abelian subgroup of $L_G$ for any subset $S$ of $R_+$ \cite{Brocker85, Hall15}. 

The proof of Theorem \ref{theo:homotopy} proceeds in four steps. 

First, we prove the following lemma on the general formulas of homotopy groups. 
\begin{lemm}
\label{lemm:homotopy2}
Let $i_n^\ast: \pi_n(H) \to \pi_n(G)$ be the induced homomorphism of the inclusion map $i: H \to G$.
The homotopy groups of the homogeneous space $\pi_i(G/H) \ (i = 1,2)$ are calculated as follows:\\
(1) For a Lie group $G$ and its connected subgroup $H$,
\begin{eqnarray}
\pi_1(G/H) = \mathrm{Coker}\{ i_1^\ast: \pi_1(H) \to \pi_1(G) \}
, \label{s4}
\end{eqnarray}
where $\mathrm{Coker}\{f: X \to Y \}$ for a homomorphism $f: X \to Y$ is defined as
\begin{eqnarray}
\mathrm{Coker}\{f: X \to Y \} := Y/\mathrm{Im}\{f: X \to Y \}
.  
\end{eqnarray}
Let $[a] \ (a \in \pi_1(G))$ and $\theta \in [0,2\pi]$ be 
a representative element of the coset space $\mathrm{Coker} \ i_1^\ast$ 
and the azimuth angle around the vortex. 
The texture $O(\theta)$ of the vortex associated with $[a]$ is given by 
\begin{eqnarray}
O(\theta) = a(\theta) \circ O_0
, \label{eq:textureofvortex}
\end{eqnarray}
where $O_0$ is the value of the order parameter at $\theta=0$ 
and $g \circ O_0 \ (g \in G)$ is the action of $g$ on $O_0$. \\
(2) For a compact Lie group $G$ and its subgroup $H$,
\begin{eqnarray}
\pi_2(G/H) = \mathrm{Ker}\{ i_1^\ast: \pi_1(H) \to \pi_1(G) \}
. \label{s5}
\end{eqnarray}
Let $\sigma$ and $\sigma_\theta \ (0 \le \theta \le \pi)$ be an element of $\mathrm{Ker} \ i_1^\ast$ 
and a continuous deformation from $\sigma_{\theta=0} = \sigma$ to $\sigma_{\theta=\pi} = e$ (trivial loop). 
The texture $O(\theta,\phi)$ of the point defect associated with $\sigma$ is given by 
\begin{eqnarray}
O(\theta, \phi) = \sigma_\theta(\phi) \circ O_0
, \label{eq:textureofpd}
\end{eqnarray}
where $\theta\in [0,\pi]$ and $\phi\in [0,2\pi]$ are the three-dimensional polar coordinates 
and $O_0$ is the value of the order parameter at $\theta=\phi=0$.

\end{lemm}
\textit{Proof of Lemma \ref{lemm:homotopy2}}\\
We first prove the formula
\begin{eqnarray}
\frac{\pi_n(G/H)}{\mathrm{Coker} \  i^\ast_n} = \mathrm{Ker} \  i_{n-1}^\ast \label{s6}
\end{eqnarray}
by using the following homotopy exact sequence \cite{Steenrod99}:
\begin{eqnarray}
&\xrightarrow[]{i_2^\ast}& \pi_2(G) \xrightarrow[]{p_2^\ast} \pi_2(G/H) \xrightarrow[]{\partial_2^\ast} \pi_1(H) \xrightarrow[]{i_1^\ast}  \pi_1(G) \xrightarrow[]{p_1^\ast} 
\nonumber \\
&\xrightarrow[]{p_1^\ast} & 
\pi_1(G/H) \xrightarrow[]{\partial_1^\ast} \pi_0(H) \xrightarrow[]{i_0^\ast} \pi_0(G),
\end{eqnarray}
where $p_n^\ast: \pi_n(G) \to \pi_n(G/H)$ and $\partial_n^\ast: \pi_n(G/H) \to \pi_{n-1}(G/H)$ are the induced homomorphism of the projection map $p: G \to G/H$ and the boundary map $\partial: G/H \to H$.

By using the homomorphism theorem and the exact sequence,
we obtain
\begin{eqnarray}
\frac{\pi_n(G/H)}{\mathrm{Ker} \ \partial_n^\ast} &=& \mathrm{Im} \ \partial_n^\ast
, \label{exact1}\\
\mathrm{Im} \ \partial_n^\ast &=& \mathrm{Ker} \  i_{n-1}^\ast
, \label{exact2}\\
\mathrm{Ker} \ \partial_n^\ast &=& \mathrm{Im} \  p_n^\ast = \frac{\pi_n(G)}{\mathrm{Ker} \  p_n^\ast}
= \mathrm{Coker} \  i_n^\ast
. \label{exact3}
\end{eqnarray}
Combining Eqs.~(\ref{exact1})-(\ref{exact3}), we obtain Eq.~(\ref{s6}).
For a connected subgroup $H$, we have $\pi_0(H) = 0$ and hence $\mathrm{Ker} \  i_0^\ast =  0$. 
Thus, we obtain Eq.~(\ref{s4}). 
Let $O$ and $[a] \ (a \in \pi_1(G))$ be 
an element of $\pi_1(G/H)$ and the element of $\mathrm{Coker} \  i_1^\ast$ that corresponds to $O$ in Eq.~(\ref{s4}). 
Let $O_0$ is a value of the order parameter. 
The projection $p: G \to G/H$ coincides with an action on $O_0$: 
\begin{eqnarray}
p(g) = g \circ O_0
. 
\end{eqnarray}
In fact, any element $h$ in $H$ is mapped by $p$ to $O_0$, $p(h) = O_0$ 
because an element of $H$ acts on $O_0$ trivially. 
Since $p_1^\ast: \pi_1(G) \to \pi_1(G/H)$ is the induced homomorphism of the projection map $p: G \to G/H$, 
the element $O$ of $\pi_1(G/H)$ corresponding to the element of $[a] \ (a \in \pi_1(G))$ is obtained 
by the projection $p$ as follows \cite{Steenrod99}: 
\begin{eqnarray}
O(\theta) = p[a(\theta)] = a(\theta) \circ O_0
. 
\end{eqnarray}
Thus, we obtain Eq.~(\ref{eq:textureofvortex}). 
Since $\pi_2(G) =  0$ for a compact Lie group $G$ \cite{Brocker85}, we obtain $\mathrm{Coker} \ i_2^\ast = 0$ and hence Eq.~(\ref{s5}). 
Let $\sigma$ and $\sigma_\theta \ (0 \le \theta \le \pi)$ be 
the element of $\mathrm{Ker} \  i_1^\ast$ 
and the path of a continuous deformation from $\sigma_{\theta=0} = \sigma$ to the trivial loop $\sigma_{\theta=\pi} = e$. 
We can obtain the element $O$ of $\pi_2(G/H)$ that corresponds to $\sigma$ in Eq.~(\ref{s5}) as follows: 
\begin{eqnarray}
O(\theta, \phi) = \sigma_\theta(\phi) \circ O_0
, \label{eq:inverseimage}
\end{eqnarray}
where $\theta \in [0,\pi]$ and $\phi \in [0,2\pi]$ are the three-dimensional polar coordinates 
and $O_0$ is the value of the order parameter at $\theta=\phi=0$. 
Thus, we obtain Eq.~(\ref{eq:textureofpd}), 
which completes the proof of Lemma \ref{lemm:homotopy2}. 

Second, we prove Eq.~(\ref{hom2}) on the second homotopy group. 
Let $L_H$ be an integral lattice of $H$. 
We define the co-root lattice $L_{R,G}$ for $G$ by 
\begin{eqnarray}
L_{R, G} := \mr{Span}_{\mathbb{Z}}\left\{\left.
\frac{2\Bal}{(\Bal, \Bal)} \right| \Bal \in R_+
\right\}
.
\end{eqnarray}
It is known that the co-root lattice $L_{R,G}$ is a subgroup of the integral lattice $L_G$ 
and that the coset space $L_G/L_{R,G}$ is equivalent to the first homotopy group $\pi_1(G)$ \cite{Brocker85, Hall15}: 
\begin{eqnarray}
\pi_1(G) = L_G/L_{R,G}
. \label{eq:firsthg}
\end{eqnarray}
Also, we define $L_{R,H}$ as the co-root lattice of the subgroup $H$. 
For $\mu$-SB, $L_H$ is a subgroup of $L_G$ and 
$L_{R,H}$ is a subgroup of $L_{R,G}$. 
Writing an element of $\pi_1(H) = L_H/L_{R,H}$ as $\bm{t} + L_{R,H} (\bm{t} \in L_H)$, 
the inclusion map $i_1^\ast: \pi_1(H) \to \pi_1(G)$ satisfies 
\begin{eqnarray}
i_1^\ast(\bm{t} + L_{R,H}) = \bm{t} + L_{R,G}
. \label{sA}
\end{eqnarray}
From Eq.~(\ref{s5}), we have
\begin{eqnarray}
\pi_2(G/H)
 &=& \mr{Ker}\tbr{i_1^\ast: \pi_1(H) \to \pi_1(G)}
\nonumber \\
 &=& \tbr{\bm{t} + L_{R,H} | \bm{t} \in L_H, i_1^\ast(\bm{t} + L_{R,H}) \in L_{R,G}}
\nonumber \\
\nonumber \\
 &=& \tbr{\bm{t} + L_{R,H} | \bm{t} \in L_H, \bm{t} \in L_{R,G}}
\nonumber \\
 &=& (L_H \cap L_{R,G})/L_{R,H}
. 
\end{eqnarray}
For $\mu$-SB with $\hmu$, 
$L_{R,G}$ and $L_{R,H}$ can be rewritten from Theorem \ref{theo:unbroken} as 
\begin{eqnarray}
L_{R,G} &=& L_R(R_+)
,\\
L_{R,H} &=& L_R(R_+ \backslash R_H)
. 
\end{eqnarray}
Therefore, we obtain the third row of Eq.~(\ref{hom2}). 
Since $L_R(R_+)$ is a subgroup of $L_G$ and all of the Cartan generators are not broken ones for DLRO, 
we have $L_H \cap L_{R,G} = L_{R,G} = L_R(R_+)$. 
Therefore, we obtain the first row of Eq.~(\ref{hom2}). 
For $\mu$-SB with $\zmu$, 
$L_{R,G}$ and $L_{R,H}$ can be rewritten from Theorem \ref{theo:unbroken} as 
\begin{eqnarray}
L_{R,G} &=& L_R(R_+)
,\\
L_{R,H} &=& 
\begin{cases}
L_R(R_+ \backslash R_0^{\mr{D}})
&\ \mr{for} \ \mr{DLRO \ and} \ \zmu; \\
L_R(R_+ \backslash R_0^{\mr{OD}})
&\ \mr{for} \ \mr{ODLRO \ and} \ \zmu, 
\end{cases}
\nonumber\\
\end{eqnarray}
and $L_H$ is a subgroup of $L_{R,G}$. 
Thus, we obtain the second and fourth rows of Eq.~(\ref{hom2}).
The texture of the point defect in Eq.~(\ref{eq:texpointdefect}) is obtained by using Eq.~(\ref{eq:textureofpd}). 
Let $\sigma$ be an element of $\mr{Ker} \ i_1^\ast$ with co-root vector $2\Bal/(\Bal, \Bal)$. 
Here, $\sigma$ represents the loop on $H$ defined by 
\begin{eqnarray}
\sigma(\phi) = \exp\left[ i\phi \frac{2H_{\Bal}}{(\Bal, \Bal)} \right]
 \ \mr{for} \ \phi \in [0,2\pi]
. 
\end{eqnarray}
Since $\sigma \in \mr{Ker} \ i_1^\ast$, there exists a continuous deformation from $\sigma$ to the trivial loop. 
In fact, 
\begin{eqnarray}
\sigma_\theta(\phi) &=& \mr{e}^{- i\theta \frac{E_{\Bal}^I}{(\Bal, \Bal)} }\mr{e}^{ i\phi \frac{H_{\Bal}}{(\Bal, \Bal)} }\mr{e}^{i\theta \frac{E_{\Bal}^I}{(\Bal, \Bal)} }\mr{e}^{ i\phi \frac{H_{\Bal}}{(\Bal, \Bal)} }
\nonumber\\
&& \ \mr{for} \ (\theta,\phi) \in [0,\pi] \times [0,2\pi]
\end{eqnarray}
describes a continuous deformation from 
$\sigma_{\theta=0} = \sigma$ to the trivial loop $\sigma_{\theta=\pi} = e$: 
\begin{eqnarray}
\sigma_{\theta=0}(\phi) &=& \mr{e}^{ i\phi \frac{H_{\Bal}}{(\Bal, \Bal)} } \mr{e}^{ i\phi \frac{H_{\Bal}}{(\Bal, \Bal)} } = \sigma(\phi)
, \\
\sigma_{\theta=\pi}(\phi) &=& 
\mr{e}^{- i\pi \frac{E_{\Bal}^I}{(\Bal, \Bal)} } \mr{e}^{ i\phi \frac{H_{\Bal}}{(\Bal, \Bal)} } \mr{e}^{i\pi \frac{E_{\Bal}^I}{(\Bal, \Bal)} } \mr{e}^{ i\phi \frac{H_{\Bal}}{(\Bal, \Bal)} }
\nonumber\\
&=&
\mr{e}^{- i\phi \frac{H_{\Bal}}{(\Bal, \Bal)} } \mr{e}^{ i\phi \frac{H_{\Bal}}{(\Bal, \Bal)} } = e
. 
\end{eqnarray}
Here, we use 
\begin{eqnarray}
\mr{e}^{- i\pi \frac{E_{\Bal}^I}{(\Bal, \Bal)} } H_{\Bal} \mr{e}^{i\pi \frac{E_{\Bal}^I}{(\Bal, \Bal)} } = - H_{\Bal}
. 
\end{eqnarray}
From Eq.~(\ref{eq:textureofpd}), 
the element $O^\prime \in \pi_2(G/H)$ corresponding to $\sigma$ 
can be obtained by acting $\sigma_\theta(\phi)$ on $O_0$: 
\begin{eqnarray}
O^\prime(\theta, \phi) &=& 
\mr{e}^{- i\theta \frac{E_{\Bal}^I}{(\Bal, \Bal)} }\mr{e}^{ i\phi \frac{H_{\Bal}}{(\Bal, \Bal)} }\mr{e}^{i\theta \frac{E_{\Bal}^I}{(\Bal, \Bal)} }\mr{e}^{ i\phi \frac{H_{\Bal}}{(\Bal, \Bal)} } \circ O_0
\nonumber\\
&=& 
\mr{e}^{- i\theta \frac{E_{\Bal}^I}{(\Bal, \Bal)} }\mr{e}^{ i\phi \frac{H_{\Bal}}{(\Bal, \Bal)} }\mr{e}^{i\theta \frac{E_{\Bal}^I}{(\Bal, \Bal)} } \circ O_0
, 
\end{eqnarray}
where we use the fact that $H_{\Bal}$ is an unbroken generator. 
Through the continuous deformation 
\begin{eqnarray}
O^\prime_u(\theta, \phi) = \mr{e}^{i u \theta \frac{E_{\Bal}^I}{(\Bal, \Bal)} }
O^\prime(\theta, \phi)
 \ \mr{for} \ 0 \le u \le 1
, 
\end{eqnarray}
$O^\prime_u$ is deformed from $O_{u=0}^\prime = O^\prime$ to $O_{u=1} = O$. 
Thus, Eq.~(\ref{eq:texpointdefect}) is obtained. 

Third, we show $\mr{Coker} \ i_1^\ast = 0 $ for DLRO. 
From Eqs.~(\ref{s4}) and (\ref{sA}), we have
\begin{eqnarray}
\mr{Im} \ i_1^\ast  &=& \tbr{i_1^\ast(\bm{t} + L_{R,H}) | \bm{t} \in L_H} 
\nonumber \\
 &=& \tbr{\bm{t} + L_{R,G} | \bm{t} \in L_H} 
\nonumber \\
 &=& \mr{Span}_{\mbZ}\tbr{L_H \cup L_{R,G}}/L_{R,G}
,
\end{eqnarray}
and hence we obatin from Eq.~(\ref{eq:firsthg}) 
\begin{eqnarray}
\mr{Coker} \ i_1^\ast &=& \frac{L_G}{\mr{Span}_{\mbZ}\tbr{L_H \cup L_{R,G}}}
. 
\end{eqnarray}
For a $\mu$-SB with DLRO, no Cartan generators are broken ones. 
Therefore, we obtain 
\begin{eqnarray}
L_G &=& L_H = L_{R,G} 
, \\
\mr{Coker} \ i_1^\ast &=& \frac{L_G}{\mr{Span}_{\mbZ}\tbr{L_G \cup L_G}} = 0
.
\end{eqnarray}

Finally, we calculate $\mr{Coker} \ i_1^\ast$ for a $\mu$-SB with ODLRO. 
For a $\mu$-SB with ODLRO, 
the Cartan subalgebla $\Gg_C$ of $G$ is included in $\Gh$ except for one direction $\mr{Span}_{\mbZ}\tbr{I} \simeq \mbZ$ 
from Theorem \ref{theo:unbroken}. 
Thus, $\mr{Coker} \ i_1^\ast$ can be written as 
\begin{eqnarray}
\mr{Coker} \ i_1^\ast
&=&
\frac
{\mr{Span}_{\mbZ}\tbr{I}}
{\mr{Span}_{\mbZ}\tbr{I} \cap \mr{Span}_{\mbZ}\tbr{L_H \cup L_{R,G}}}
. \label{sB}
\nonumber\\
\end{eqnarray}
Let $a_n \ (n \in \mathbb{Z})$ be an element of $\mr{Span}_{\mbZ}\tbr{I}$ in the numerator of Eq.~(\ref{sB}). 
Since the action of $a_n(\theta)$ on the order parameter 
represents the rotation of the phase $2n\pi$ around the vortex, 
Eq.~(\ref{eq:textureofvortex}) 
reduces to Eq.~(\ref{eq:texvortex}). 

For the ODLRO with $\zmu$, 
$L_H$ is a subgroup of $L_{R,G}$ from Theorem \ref{theo:unbroken}. 
We obtain from Theorem \ref{theo:homotopy} 
\begin{eqnarray}
\mr{Span}_{\mbZ}\tbr{I} \cap \mr{Span}_{\mbZ}(L_H \cap L_{R,G}) 
&=& \mr{Span}_{\mbZ}\tbr{I} \cap L_{R,G} 
\nonumber\\
&=& \emptyset
, \\
\mr{Coker} \ i_1^\ast &=& \mr{Span}_{\mbZ}\tbr{I}
. 
\end{eqnarray}
Therefore, we obtain $\mr{Coker} \ i_1^\ast = \mbZ$ from Eq.~(\ref{sB}). 
For the ODLRO with $\hmu$, 
$\mr{Coker} \ i_1^\ast$ is a subgroup of $\mbZ$. 
To show that $\mr{Coker} \ i_1^\ast$ is a finite group, 
it is sufficient to show 
\begin{eqnarray}
n I \in \mr{Span}_{\mbZ}(L_H \cup L_{R,G})
\ \mr{for} \ \exists n \in \mbZ
. \label{2ssb3}
\end{eqnarray}
Let $\bar{r}$ be the rank of the Lie algebra $\bar{\Gg}$ and 
let $C = \tbr{C_{ij}}_{i,j =1}^{\bar{r}}, \tbr{\Bal^{(j)}}_{j=1}^{\bar{r}}$, and $\tbr{m_i}_{i=1}^{\bar{r}}$ be 
the Cartan matrix of $\bar{\Gg}$, a set of simple roots of $\bar{\Gg}$, and the Dynkin index of $\hmu$, respectively. 
The generator $I$ can be written as
\begin{eqnarray}
I &=& \frac{1}{|\hmu|^2} \left( |\hmu|^2 I - H_{\hmu}\right) 
\nonumber \\
&&+ \sum_{j=1}^{\bar{r}} 
\left[ \frac{1}{|\hmu|^2} \sum_{i=1}^{\bar{r}} m_i C_{ij} \frac{(\Bal^{(j)}, \Bal^{(j)})}{2} \right] 
\frac{2 H_{\Bal^{(j)}}}{(\Bal^{(j)}, \Bal^{(j)})}
. 
\nonumber \\
\end{eqnarray}
Since $|\hmu|^2 I - H_{\hmu}$ and $\frac{2 H_{\Bal^{(j)}} }{(\Bal^{(j)}, \Bal^{(j)})}$ are the elements of $L_H \cup L_{R,G}$ 
and the coefficients are all rational numbers, 
the greatest common divisor of the denominator of the coefficients satisfies Eq.~(\ref{2ssb3}). 
The positive integer $l$ in Eq.~(\ref{hom1}) is determined to be the minimum number that satisfies Eq.~(\ref{2ssb3}), 
which completes the proof of Theorem \ref{theo:homotopy}. 

\bibliography{citation}


\end{document}